\documentclass[aps,prl,showpacs,twocolumn,superscriptaddress,amsmath,amssymb,floatfix,longbibliography,footinbib]{revtex4-1}
\usepackage{graphicx}
\usepackage{float}
\usepackage{latexsym,amsmath,amssymb,bm,euscript}
\usepackage{color}
\usepackage{wasysym}
\usepackage{epstopdf}
\usepackage[colorlinks=true,linkcolor=blue,citecolor=blue]{hyperref}
\usepackage{type1cm}
\usepackage{appendix}
\usepackage{mathtools}
\usepackage{booktabs}
\usepackage[dvipsnames]{xcolor}
\usepackage{bbold}
\usepackage{amssymb}
\usepackage{stackengine}
\usepackage{scalerel}
\usepackage{xcolor}
\usepackage{graphicx}
\usepackage{bbm}

\usepackage{braket}

\usepackage{tikz}
\usetikzlibrary{topaths}

\newcommand\sblacksquare[1][.5]{\mathbin{\vcenter{\hbox{\scalebox{#1}{$\blacksquare$}}}}}

\newcommand{\ketbra}[2]{\mathinner{|{#1}\rangle\! \langle{#2}|}}

\usepackage{xcolor}
    \definecolor{myred}{HTML}{E07A60}
    \definecolor{myblue}{HTML}{687AB1}
    \definecolor{mygreen}{HTML}{66b063}
    \definecolor{myviolet}{HTML}{8744f9}

\newcommand{\tcb}{\textcolor{blue}}

\usepackage{soul}

\newcommand*{\tran}{{\mkern-1.5mu\mathsf{T}}}

\usepackage[normalem]{ulem}

\begin{document}

\title{Density-matrix renormalization group algorithm for non-Hermitian systems}

\author{Peigeng Zhong}
\affiliation{Beijing Computational Science Research Center, Beijing 100084, China}

\author{Wei Pan}
\affiliation{Beijing Computational Science Research Center, Beijing 100084, China}

\author{Haiqing Lin}
\email[]{haiqing0@csrc.ac.cn}
\affiliation{Beijing Computational Science Research Center, Beijing 100084, China}
\affiliation{School of Physics \& Institute for Advanced Studies of Physics, Zhejiang University, Hangzhou, 310058, China}

\author{Xiaoqun Wang}
\email[]{xiaoqunwang@zju.edu.cn}
\affiliation{School of Physics \& Institute for Advanced Studies of Physics, Zhejiang University, Hangzhou, 310058, China}

\author{Shijie Hu}
\email[]{shijiehu@csrc.ac.cn}
\affiliation{Beijing Computational Science Research Center, Beijing 100084, China}
\affiliation{Department of Physics, Beijing Normal University, Beijing, 100875, China}

\begin{abstract}
A biorthonormal-block density-matrix renormalization group algorithm is proposed to accurately compute properties of large-scale non-Hermitian many-body systems, in which a renormalized-space partition of the non-Hermitian reduced density matrix is implemented to fulfill the prerequisite for the biorthonormality of the renormalization group (RG) transformation and to optimize the construction of saved Hilbert spaces. A redundancy in saved spaces of the reduced density matrix is exploited to reduce a condition number resulting from the non-unitarity of the left and right transformation matrices, in order to ensure the numerical stability of the RG procedure. The algorithm is successfully applied to an interacting fermionic Su-Schrieffer-Heeger model with nonreciprocal hoppings and staggered complex chemical potential, exhibiting novel many-body phenomena.
\end{abstract}

\maketitle

Non-Hermitian quantum systems in recent years have become of great interest in the exploration of the intriguing biorthogonal physics associated with nontrivial topology~\cite{Bergholtz_2021}, exceptional points (EPs)~\cite{ElGanainy_2018}, and non-Hermitian skin effects~\cite{Zhang_2022c}. These phenomena are observable in photonic quantum walks~\cite{Xiao_2020}, ultracold atomic gases~\cite{Gou_2020}, and other interdisciplinary studies~\cite{Gao_2015, Zhang_2021, Wang_2022}. More lately, non-Hermitian many-body effects have been increasingly addressed for spin liquids~\cite{Yang_2021b}, topological states~\cite{Zhang_2020, Zhou_2020c, Liu_2020c, Lee_2021, Faugno_2022, Kawabata_2022, Zhen_2022, Chen_2023, Yoshida_2023}, and fractional quantum Hall states~\cite{Yoshida_2020a}, and antiferromagnetic ordering~\cite{Yu_2023}. However, most efforts so far have mainly been cast into a few integrable models~\cite{Shackleton_2020, Nakagawa_2021} and special limits~\cite{Yamamoto_2022a}, demanding numerical tools for reliably simulating large-scale general non-Hermitian systems described by Liouvillians or Hamiltonians~\cite{Sa_2020}.

\begin{figure}[b]
\includegraphics[width=\columnwidth]{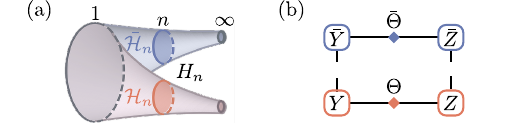}
\caption{\label{fig:fig1}(Color online) (a) A schematic RG procedure of the Hilbert spaces $\mathcal{H}_n$ (red ellipsoid) and $\bar{\mathcal{H}}_n$ (blue ellipsoid), sharing an initialization $\mathcal{H}_1 = \bar{\mathcal{H}}_1$ (leftmost and dark ellipsoid), is directed towards a fixed point of $n=\infty$ (two smaller ellipsoids). The Hamiltonian $H_n$ maps $\mathcal{H}_n$ to $\bar{\mathcal{H}}_n$ at step $n$. (b) A sketch of the decompositions of the left and right eigenstates for $H$ with respect to two semi-chains (see text).}
\end{figure}

The renormalization group (RG) theory has played an important role in determining the correlation effects in many-body systems~\cite{Ma_1973, Wilson_1975, Wilson_1983, Shankar_1994, Fisher_1998}. Typically, it is crucial to identify a way to transform a bare Hamiltonian into an effective one for low-energy physics, through iterative RG operations that successively renormalize the Hilbert space. Figure~\ref{fig:fig1}(a) depicts a general RG procedure for a non-Hermitian system, where the right bases that span the Hilbert space $\mathcal{H}_n$ are biorthonormal to the left bases that span the dual space $\bar{\mathcal{H}}_n$ in the $n$-th RG step. Usually, $\mathcal{H}_n \neq \bar{\mathcal{H}}_n$ except for $n=1$, at which both $\mathcal{H}_1$ and $\bar{\mathcal{H}}_1$ are constructed in terms of the bare bases. Meanwhile, a similarity transformation, defined by $\mathbb{W}_n [H_n] \equiv \bar{W}_n H_n W_n$, is adopted to renormalize the Hamiltonian $H_n$, preserving its spectrum. When truncation is made for both $\mathcal{H}_n$ and $\bar{\mathcal{H}}_n$, it is found that $\bar{W}_n W_n = \mathbb{1}$ but $\bar{W}_n \neq W_n^{-1}$. In a Hermitian case, one finds that $\mathcal{H}_n = \bar{\mathcal{H}}_n$, and the RG transformation $\mathbb{U}_n [H_n] \equiv U^\dag_n H_n U_n$ can be established with a unitary matrix $U_n$~\cite{Wilson_1983}. The necessity of utilizing the similarity transformation for the non-Hermitian case can be demonstrated in a quasi-Hermitian system, where an invertible matrix $\nu_n$ is used to find a Hermitian counterpart $H^{(\text{h})}_n$ via a transformation $H^{(\text{h})}_n = \nu_n H_n \nu^{-1}_n$~\cite{Mostafazadeh_2002}. Subsequently, a relation $U_n = \nu_n W_n \nu^{-1}_{n+1}$ is established between two distinct RG procedures. However, navigating correct RG transformations for general non-Hermitian many-body systems poses severe challenges~\cite{Yamamoto_2022a}.

The Density-matrix renormalization group (DMRG) method has achieved great success in studying low-energy properties of (quasi-)one-dimensional Hermitian Hamiltonians~\cite{White_1992, Peschel_1999, Schollwock_2005} and in extending to two dimensions~\cite{White_2007, Luo_2017, Hu_2019, Xiang_2023}. However, it has so far failed to be applied reliably in exploring the properties of interacting non-Hermitian Hamiltonians~\cite{Henkel_2001, Affleck_2004, Carlon_2001b, Hieida_1998, Temme_2010, Johnson_2015, Zhang_2020, Yamamoto_2022a}, because the similarity transformation cannot be properly constructed with a non-Hermitian reduced density matrix in the standard DMRG framework~\cite{Wang_1997, Xiang_1998, Shibata_2003, Affleck_2004, Chan_2005, Huang_2014}, resulting inevitably in severe numerical instability~\cite{Henkel_2001, Affleck_2004, Carlon_2001b, Hieida_1998, Temme_2010, Johnson_2015}. Since non-Hermitian physics becomes of increasing interest, it is expected that a broader DMRG algorithm is developed for reliably studying general non-Hermitian problems.

In this letter, we develop a biorthonormal-block DMRG (bbDMRG) algorithm that resolves the challenge of implementing similarity transformations, which are necessarily used to construct the saved Hilbert space and its dual.
Our algorithm introduces a renormalized-space partition of the non-Hermitian reduced density matrix and utilizes the features of redundant degrees of freedom to mitigate numerical instability, enabling highly accurate large-scale numerical simulations for interacting non-Hermitian systems.

\textit{Biorthonormal-block DMRG}.---Let us consider a chain consisting of a left semi-chain ($\triangleleft$) and a right semi-chain ($\triangleright$).
The space $\mathcal{H} = \mathcal{H}^\triangleleft \otimes \mathcal{H}^\triangleright$ and its dual space $\bar{\mathcal{H}} = \bar{\mathcal{H}}^\triangleleft \otimes \bar{\mathcal{H}}^\triangleright$ are products of the spaces for semi-chains $\mathcal{H}^{\triangleleft,\triangleright}$ and $\bar{\mathcal{H}}^{\triangleleft,\triangleright}$, respectively. The left eigenstate $\bra{\bar{\Psi}}$ and the corresponding right eigenstate $\ket{\Psi}$ for the non-Hermitian Hamiltonian $H$ satisfy $\braket{\bar{\Psi} \vert \Psi} = 1$. Notably, we employ a two-sided Krylov-Schur-restarted Arnoldi diagonalization technique~\cite{Zwaan_2017}, which shows advantages especially in systems with degenerate energy spectra (End Matter~A).

To elucidate the bbDMRG algorithm, we start by focusing on the left semi-chain $\triangleleft$ to demonstrate how RG transformations are specifically customized for non-Hermitian Hamiltonians during the chain growth and sweeping processes~\cite{SM}. At each bbDMRG step, a non-Hermitian reduced density matrix $\rho = \text{tr}_\triangleright \ketbra{\Psi} {\bar{\Psi}}$ resides in the spaces $\mathcal{H}^\triangleleft$ and $\bar{\mathcal{H}}^\triangleleft$ spanned by bases $\ket{y^0_\beta}$ and $\bra{\bar{y}^0_\beta}$, satisfying $\braket{\bar{y}^0_\beta \vert y^0_{\beta^\prime}} = \delta_{\beta,\beta^\prime}$ (End Matter~B). Indices $\beta$, $\beta^\prime = 1$, $\cdots$, $d^\triangleleft$, where $d^\triangleleft$ denotes the dimension of spaces for $\triangleleft$.
A key ingredient of each step is to partition $\rho$ into
\begin{equation}\label{eq:EqBlock}
	\rho =
	\begin{pmatrix}
		Y^{(\text{s})} & Y^{(\text{d})}
	\end{pmatrix}
	\begin{pmatrix}
		B^{(\text{s})} & \mathbb{0} \\
		\mathbb{0}  & B^{(\text{d})}
        \end{pmatrix}
        \begin{pmatrix}
		\bar{Y}^{(\text{s})} \\ \bar{Y}^{(\text{d})}
        \end{pmatrix}\, ,
\end{equation}
which defines $\rho^{(\text{s},\text{d})} = {Y}^{(\text{s},\text{d})} B^{(\text{s},\text{d})} \bar{Y}^{(\text{s},\text{d})}$ with two blocks $B^{(\text{s},\text{d})}$ such that $[\rho^{(\text{s})}, \rho^{(\text{d})}]=0$. In this \textit{renormalized-space partitioning process}, we construct saved (s) Hilbert spaces with the dimension of $m \le d^\triangleleft$, by forming the rectangular matrices $Y^{(\text{s})}$ and $\bar{Y}^{(\text{s})}$ for the RG transformation, respectively.
While the rest, i.e., $Y^{(\text{d})}$ and $\bar{Y}^{(\text{d})}$, is discarded (d) when the off-diagonal terms in $\rho$ vanish, thereby rendering $\rho^{(\text{s})}$ a low-rank approximation to $\rho$. These RG operations for $\triangleleft$ are also applicable to $\triangleright$, yielding another RG transformation matrices $Z^{(\text{s})}$ and $\bar{Z}^{(\text{s})}$ that act upon the biorthonormal bases $\ket{z^0_\beta}$ and $\bra{\bar{z}^0_\beta}$ spanning the spaces $\mathcal{H}^\triangleright$ and $\bar{\mathcal{H}}^\triangleright$, respectively. Clearly, $\bar{Y}^{(\text{s})}Y^{(\text{s})} = \bar{Z}^{(\text{s})} Z^{(\text{s})} = \mathbbm{1}$.

We note that partitioning $\rho$ into two blocks, rather than a full diagonalization or singular value decomposition (SVD) used in the traditional DMRG, results in beneficial redundancy in the construction of saved spaces. It is evident that the block-diagonal form presented in Eq.~\eqref{eq:EqBlock} remains unchanged, when applying an invertible matrix $\eta$ to saved spaces as follows:
\begin{eqnarray}
\label{eq:redundancy}
Y^{(\text{s})}_\eta = Y^{(\text{s})} \eta\, , \quad \bar{Y}^{(\text{s})}_\eta = \eta^{-1} \bar{Y}^{(\text{s})}\, .
\end{eqnarray}
This also implies that the sparsity of the block $B^{(\text{s})}_\eta = \bar{Y}^{(\text{s})}_\eta \rho Y^{(\text{s})}_\eta$ can be adjusted by selecting appropriate $\eta$.

Structurally, the bbDMRG algorithm looks for the target left and right eigenstates, represented as [Fig.~\ref{fig:fig1}(b)]
\begin{eqnarray}\label{eq:bWF}
\bra{\bar{\Psi}} \! = \! \sum_{\alpha,\alpha^\prime} \bra{\bar{y}_\alpha} \bar{\Theta}_{\alpha,\alpha^\prime} \bra{\bar{z}_{\alpha^\prime}}\, , \ \ket{\Psi} \! = \! \sum_{\alpha,\alpha^\prime} \ket{y_\alpha} \Theta_{\alpha,\alpha^\prime} \ket{z_{\alpha^\prime}}\, ,\ 
\end{eqnarray}
where $\bra{\bar{y}_\alpha} = \sum_{\beta} \bar{Y}^{(\text{s})}_{\alpha,\beta} \bra{\bar{y}^0_\beta}$, $\bra{\bar{z}_{\alpha'}} = \sum_{\beta} \bar{Z}^{(\text{s})}_{\alpha',\beta} \bra{\bar{z}^0_\beta}$, $\ket{y_\alpha} = \sum_{\beta} \ket{y^0_\beta} Y^{(\text{s})}_{\beta,\alpha}$, and $\ket{z_{\alpha'}} = \sum_{\beta} \ket{z^0_\beta} Z^{(\text{s})}_{\beta,\alpha'}$ give biorthonormal transformed bases. Bond indices $\alpha$ and $\alpha^\prime$ range from $1$ to $m$.
Unlike the biorthonormal matrix-product-state representation designed for Perron states~\cite{Huang_2011}, Eq.~\eqref{eq:bWF} provides the \textit{biorthonormal-block representation} where matrices $\bar{\Theta}$ and $\Theta$ are non-diagonal. Readily, $B^{(\text{s})} = \Theta \bar{\Theta}$. The multi-step decomposition and the bbDMRG flow are shown in End Matter~B.

Prior to constructing $B^{(\text{s})}$, it is necessary to establish a \textit{truncation criterion} in terms of the measurement error $\varepsilon = \vert \text{tr} [(\rho - \rho^{(\text{s})})O] \vert = \vert \text{tr} (\rho^{(\text{d})}O) \vert$ for the expectation value of a physical observable $O$~\cite{White_1992, Schollwock_2005}. An upper bound (UB) for $\varepsilon$ can be found \textit{via} von Neumann's trace inequality~\cite{Mirsky_1975}, which gives rise to $\varepsilon \leq  \sum_\alpha \lambda^{\rho^{(\text{d})}}_\alpha \lambda^O_\alpha \equiv \varepsilon_1$, where $\lambda^{\rho^{(\text{d})}}_\alpha$ and $\lambda^O_\alpha$ correspond to the singular values of $\rho^{(\text{d})}$ and $O$ sorted in descending order. By introducing $a = \sum_\alpha \lambda^O_\alpha$, a larger UB is given as $\varepsilon \le\varepsilon_1\leq \varepsilon_2 = a \Vert \rho^{(\text{d})} \Vert_2$ with $\Vert \rho^{(\text{d})} \Vert_2$ being the $2$-norm of $\rho^{(\text{d})}$. According to the Eckart-Young-Mirsky theorem widely used  in tensor algorithms~\cite{Cirac_2021a}, $\Vert \rho^{(\text{d})} \Vert_2$ has a minimum value of $\lambda^\rho_{m+1}$, where $\lambda^\rho_\alpha$ is the $\alpha$-th largest singular value of $\rho$. Consequently, one formally obtains $\sup (\varepsilon_2) = a \lambda^\rho_{m+1}$ as the least UB, consistent with that for the Hermitian case. However, it is an unresolved and complex non-convex optimization problem in math to minimize $\varepsilon_2$, while simultaneously satisfying the non-unitary RG requirement for spanning $\mathcal{H}$ and $\bar{\mathcal{H}}$, in order to achieve the best \textit{structured} low-rank approximation~\cite{Markovsky_2008}.

To deal with this issue, we fortunately find a detour by sorting the spectrum of the eigenvalues $\zeta_\alpha$ of $\rho$ in descending order of the non-negative weight $w_\alpha = \vert \zeta_\alpha \vert$ and retaining spaces spanned by the selected pairs of $\bra{\bar{y}_\alpha}$ and $\ket{y_\alpha}$. Typically, these bases are the linear combinations of the first $m$ eigenvectors of $\rho$. As a result, the strict condition on $\epsilon_2$ can be relaxed by the introduction of a larger UB $\varepsilon_3 = a \kappa w_{m+1} \ge \varepsilon_2$ after employing the Cauchy-Schwarz inequality~\cite{SM}, where $\kappa = \Vert{Y} \Vert_2 \Vert {\bar{Y}} \Vert_2$ is a condition number. In the optimal selection, both $\kappa$ and $w_{m+1}$ are supposed to be minimized as much as possible in the implementation of $Y^{(s)}$ and $\bar{Y}^{(s)}$. Specifically, $\kappa$ measures the deviation strength of the non-Hermitian $\rho$ from the best approximation of all possible Hermitian ones to some extent. For the Hermitian case, $\kappa = 1$ and $w_\alpha =\lambda^\rho_\alpha$, resulting in $\varepsilon_3 = \varepsilon_2$ as anticipated~\cite{White_1992}. In the non-Hermitian case, where $\kappa$ can be much larger than $1$, the partition of $\rho$ following Eq.~\eqref{eq:EqBlock} can be executed such that ensures $w_m > w_{m+1}$. In this case, the singular value and weight spectra of $\rho$, used in the normal and structured low-rank approximations, respectively, have a consistent profile with a bounded discrepancy of $\left\vert \lambda^\rho_\alpha - w_\alpha \right\vert \le \Vert {T} \Vert_2$, and $T$ denotes the strictly upper triangular matrix in the Schur decomposition of $\rho$~\cite{Zhang_2017a}.
Thus, $\varepsilon_3$ effectively stands for UB of the measurement error $\varepsilon$ in bbDMRG, while $\varepsilon_t = 1 - \sum^m_{\alpha=1} w_\alpha$ is defined as the truncation error.

When $\kappa$ is large, it is impossible to construct ${Y}^{({\text s})}$, $\bar{Y}^{({\text s})}$, and $B^{({\text s})}$ directly by the full diagonalization of a non-Hermitian $\rho$ due to the significant numerical errors that arise during the calculations. To confront this challenge, we find a two-step approach to achieve a numerically stable partition of $\rho$ as described in Eq.~\eqref{eq:EqBlock}. First, we convert $\rho$ into its upper triangular form consisting of matrices $A$, $C$ and $D$, by using a Schur decomposition~\cite{Zhang_2017a}
\begin{eqnarray}\label{Schur}
	\rho = S
	\begin{pmatrix} A & D \\  \mathbb{0} & C \end{pmatrix}
	S^\dag\, ,
\end{eqnarray}
with a unitary matrix $S = (S^{(\text{s})}\ S^{(\text{d})})$. It is remarkable that diagonal elements of $A$ and $C$ are eigenvalues $\zeta_\alpha$ exactly. In this step, we sort $\zeta_\alpha$ readily in descending order of $w_\alpha$ and are allowed to separate the full space into the saved and discarded ones. 
Secondly, using the Bartels-Stewart algorithm~\cite{Bartels_1972}, we find a matrix $X$ to eliminate the matrix $D$ in Eq.~\eqref{Schur} by solving the Sylvester equation $A X - X C = D$. Lastly, in terms of $A$, $C$, $S$ and $X$, we have $Y^{(\text{s})}=S^{(\text{s})}$, $Y^{(\text{d})}=S^{(\text{d})}-S^{(\text{s})}X$, $\bar{Y}^{(\text{s})}=S^{(\text{s})\dag}+XS^{(\text{d})\dag}$, $\bar{Y}^{(\text{d})}=S^{(\text{d})\dag}$, $B^{(\text{s})}=A$, and $B^{(\text{d})}=C$~\cite{Bartels_1972}, yielding the block diagonal form~\eqref{eq:EqBlock}. The two-step approach may be replaced by other potential methods that meet the truncation criterion stated earlier. Notably, $\kappa \gg 1$ directly results in more severe numerical instability with larger error $\varepsilon$, which is the inherent obstacle to earlier DMRG exploration~\cite{Yamamoto_2022a, Zhang_2020}. The redundancy in the construction of saved spaces, as shown in Eq.~\eqref{eq:redundancy}, allows us to obtain appropriate $\eta$ for effectively reducing $\kappa$ by various skills over $Y^{(\text{s})}$ and $\bar{Y}^{(\text{s})}$ (End Matter~C). And rescaling $H^{\triangleleft,\triangleright}$ often substantially enhances the precision of bbDMRG results~\cite{Honecker_2011, Hu_2011}.

\begin{figure}[t]
\includegraphics[width=\columnwidth]{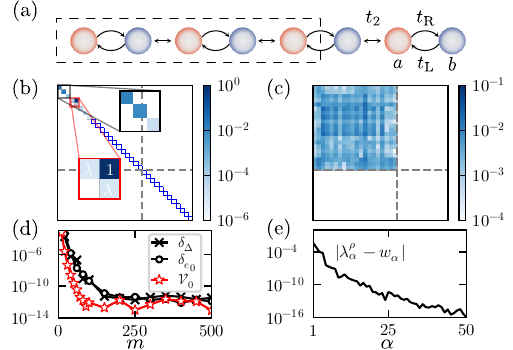}
\caption{\label{fig:fig2}
(Color online) Tests on the SSH model~\eqref{eq:HamSSH}. When a chain of $N=4$ unit cells is segmented into two semi-chains of $5$ left and $3$ right sites as shown in (a), block-diagonal forms (b) and (c) of $\rho$ are plotted for $t_1 = 1.5$, $t_2 \approx 0.96$~\cite{Note1} and $\gamma = V = 2$. (b) GEVD-based Jordan normal form~\cite{SM} consisting of ordinary eigenvalues (e.g. {\color{myblue}$\sblacksquare[0.8]$} in the black box), a block of $2 \times 2$ (red box), and null spaces ({\color{blue}$\Square$}). (c) A $2 \times 2$ block diagonal form through a two-step approach. (d) Absolute errors $\delta_{e_0}$ and $\delta_\Delta$ for the ground-state energy $e_0$ and the gap $\Delta$, respectively, as well as the ground-state full variance $\mathcal{V}_0$, versus the dimension $m$ at $N=50$, $t_1 = 0.7$ and $V=0$. (e) $\lvert \lambda^\rho_\alpha - w_\alpha \rvert$ at $N=50$, $t_1=0.7$, $V=5$ and $m=100$.}
\end{figure}

\textit{Efficiency of the algorithm}.---We now turn to applying bbDMRG to an interacting fermionic Su-Schrieffer-Heeger (SSH) model [Fig.~\ref{fig:fig2}(a)] on a duplex lattice of $N$ unit cells under open boundary conditions (OBC) as described by 
\begin{eqnarray}\label{eq:HamSSH}
\begin{split}
H_\text{SSH} &= \sum^N_{\ell=1} \left( t_\text{L} c^\dag_{\ell,a} c^{\phantom{\dag}}_{\ell,b} + t_\text{R} c^\dag_{\ell,b} c^{\phantom{\dag}}_{\ell,a} + V n^{\phantom{\dag}}_{\ell,a} n^{\phantom{\dag}}_{\ell,b} \right)\\
  &+ \sum^{N-1}_{\ell=1} \left[ t_2 (c^\dag_{\ell,b} c^{\phantom{\dag}}_{\ell+1,a} + \text{h.c.} ) + V n^{\phantom{\dag}}_{\ell,b} n^{\phantom{\dag}}_{\ell+1,a} \right] \\
  &+ \sum^{N}_{\ell=1} \sqrt{2}  e^{-i \pi / 4}  u \left( n^{\phantom{\dag}}_{\ell,a} - n^{\phantom{\dag}}_{\ell,b} \right)\, ,
\end{split}
\end{eqnarray}
where $c^\dag_{\ell,\sigma}$, $c^{\phantom{\dag}}_{\ell,\sigma}$, and $n^{\phantom{\dag}}_{\ell,\sigma} = c^\dag_{\ell,\sigma} c^{\phantom{\dag}}_{\ell,\sigma}$, with $\sigma = a$ or $b$, represent creation, annihilation, and particle-number operators for the fermion, respectively. Accordingly, the position of the site-($\ell,\sigma$) is $x=2\ell - 1$ for sublattice-$a$, and $x=2\ell$ for sublattice-$b$.
Hereafter, the index ($\ell,\sigma$) is replaced with $x$ sometimes for convenience.
The particle-number operators for fermions in two semi-chains are defined as $N^\text{f}_{\triangleleft,\triangleright} = \sum_{\ell \in \triangleleft, \triangleright} (n_{\ell,a} + n_{\ell,b})$. Hopping coefficients $t_{\text{L},\text{R}}= t_1 \pm \gamma$ have a nonreciprocity $\gamma \ge 0$ and $t_1 > 0$ for odd bonds, and $t_2 = 1$ for even bonds set as the energy unit. $V \ge 0$ represents the repulsion strength between fermions at nearest-neighbor (NN) sites, while $u$ gives the strength of the staggered complex chemical potential. Unless explicitly stated, the following discussions focus on the ground state, possessing the minimal real part of the energy, at \textit{half-filling} with $\gamma = 0.1$. We also distinguish the left-right (LR) expectation value $\braket{O}_\text{lr} = \braket{\bar{\Psi} \lvert O \rvert \Psi}$ from the right-right (RR) one $\braket{O}_\text{rr} = \braket{\Psi \lvert O \rvert \Psi} / \braket{\Psi \lvert \Psi}$ for a physical observable $O$.

Figure~\ref{fig:fig2}(b) displays the Jordan normal form of $\rho$ obtained through a regular generalized eigenvalue decomposition (GEVD) with a parameter set in the $\mathcal{PT}$-broken region for $N=4$~\cite{SM}. In this case, we kept twenty generalized eigenvectors to construct RG transformation matrices. One DMRG step yields an absolute error of $\delta_{e_0}=10^{-7}$ for the ground-state energy $e_0$, with the corresponding full variance $\mathcal{V}_0=\braket{(H_\text{SSH}-e_0)^2}_\text{lr}$~\cite{Hubig_2018} on the same order of magnitude, regardless of $\varepsilon_t = 0$, while the corresponding condition number $\kappa \approx 10^7\gg 1$. On the contrary, when the renormalized-space partitioning process is implemented using the two-step approach, $\kappa$ is reduced to $1$ with obtaining $\delta_{e_0}$ and $\mathcal{V}_0$ at machine precision, and the block for saved spaces becomes dense as shown in Fig.~\ref{fig:fig2}(c). The bbDMRG calculations are also performed at $N=50$, $t_1=0.7$ and $V=5$ to compute $e_0$ and the gap $\Delta = e_1 - e_0$, where $e_1$ is the first excitation energy based on the ascending order of the real part of energy values. One can see that both $\delta_{e_0}$ and $\delta_{\Delta}$ converge to their minimum values with increasing the dimension $m$ [Fig.~\ref{fig:fig2}(d)].
Meanwhile, $\mathcal{V}_0$ remains around or much below $10^{-12}$ for large $m$, further confirming the high quality of the ground-state wave functions obtained from bbDMRG calculations.
Figure~\ref{fig:fig2}(e) further shows the discrepancy $\left\vert \lambda^\rho_\alpha - w_\alpha \right\vert$ at $N=50$, $t_1=0.7$ and $V=5$, which is bounded and also decreases rapidly with increasing $\alpha$.
More benchmarks are provided in Supplemental Material~\cite{SM}, particularly for the interacting fermionic SSH model with third-neighbor hoppings~\cite{Yao_2018}, which lacks a Hermitian counterpart even at $u=0$.
Below are two kinds of novel kink behaviors presented for non-Hermitian many-body effects \tcb{of}  the model~\eqref{eq:HamSSH}.

\begin{figure}[t]
\includegraphics[width=0.81\columnwidth]{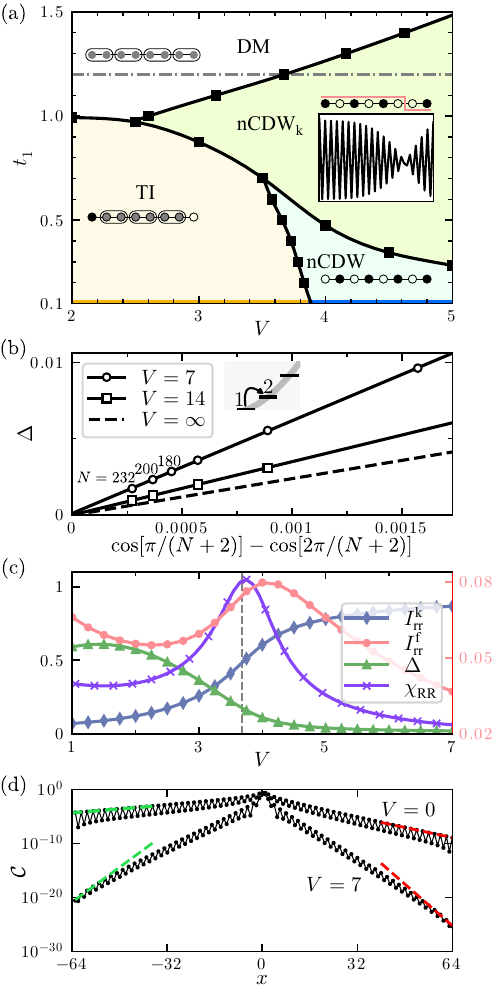}
\caption{\label{fig:fig3}(Color online) (a) The ground-state phase diagram. Black squares indicate transition points determined by peaks of the RR-fidelity susceptibility $\chi^{\phantom{\dag}}_\text{RR}$~\cite{Gu_2008, SM} with $N=48$ and $m=100$. Inset:  $\braket{n_{\ell,\sigma}}_\text{rr}$ at $t_1 = 0.7$, $V = 7$, and $N=20$. (b) The gap $\Delta$ versus $N$, vanishing as $N\rightarrow\infty$. Inset: The kink transitions from the bottom level-$1$ at the momentum $p=\pi - p_0$ to the higher level-$2$ at $p=\pi - 2p_0$ with $p_0=\pi/(N+2)$, in the band $2\sqrt{t_1^2-\gamma^2} \cos p - 2 [(t^2_1 - \gamma^2) / V] \cos (2 p)$ (grey curve) from the effective Hatano-Nelson model~\cite{Hatano_1996, SM}. 
(c) $I^{\text{k},\text{f}}_\text{rr}$, $\Delta$ and $\chi^{\phantom{\dag}}_\text{RR}$ as a function of $V$ for $N=64$. The dashed line indicates the transition point $V_c \approx 3.6$ determined in (a).
(d) Single-particle correlation function $\mathcal{C}(x)$ for $N=64$. The bipartite structure leads to even-odd oscillations with $x$.  As $\vert x\vert$ increases, $\mathcal{C}(x)$ shows distinct exponential-decay rates in the positive and negative $x$-directions, highlighted by dashed lines.
For (b)-(d), $t_1=1.2$ (dash-dotted line in (a)).}
\end{figure}

\textit{Skin effects of a kink at $u=0$}.---Figure~\ref{fig:fig3}(a) presents the ground-state phase diagram for a $\mathcal{PT}$-unbroken region of $t_1 \ge \gamma$~\cite{Alsallom_2022}. While a topological insulator (TI), a dimerized phase (DM), and a normal EP of $t_1 = \gamma$ have been explored intensively for $V=0$~\cite{Rudner_2009}, we find a non-Hermitian charge density wave (nCDW), an nCDW with a kink (nCDW\textsubscript{k}), and a CDW-EP line, for sufficiently large $V$. The nCDW phase involves two-fold degenerate ground states in which $n_{\ell,\sigma}$ oscillates on two sublattices, resulting in two different configurations, nCDW-$1$: ($1_{1,a}$, $0_{1,b}$, $\cdots$, $0_{N,b}$), and nCDW-$2$: ($0_{1,a}$, $\cdots$, $0_{N,a}$, $1_{N,b}$). The CDW-EP line at $t_1 = \gamma$ arises from the exclusion principle of fermions on two nearest-neighbor sites, rather than from skin effects in the normal EP, and cannot be simply explained in the context of the generalized Brillouin zone~\cite{Yao_2018}.

In the  nCDW\textsubscript{k} phase, a pair of holes emerges to form a \textit{kink}, separating a left nCDW-$1$ from a right nCDW-$2$, which results in a novel strong skin effect as shown by $\braket{n_{\ell,\sigma}}_\text{rr}$ and the position of the kink [inset of Fig.~\ref{fig:fig3}(a)]. The motion of the kink yields a gap $\Delta = \chi [ \cos(\pi/(N+2)) - \cos(2\pi/(N+2)) ]$ for the lowest excitation [Fig.~\ref{fig:fig3}(b)]. As both $V$ and $N$ approach $+\infty$, $\chi$ gradually converges to $\chi_0 = 2 \sqrt{t^2_1 - \gamma^2}$. For finite $V$, $\chi \approx \chi_0 + 8 (t^2_1 - \gamma^2) / V > \chi_0$ reflects the contribution of higher-order processes~\cite{SM}.

To interpret the nature of the nCDW\textsubscript{k} phase, we calculate the imbalance $I^{\text{f},\text{k}}_\text{rr} =$$\lvert \braket{N^{\text{f},\text{k}}_\triangleleft - N^{\text{f},\text{k}}_\triangleright}_\text{rr} \rvert$ of the fermion number and the kink number $N^\text{k}_{\triangleleft,\triangleright} = \sum_{\ell \in \triangleleft,\triangleright} [(1 - n_{\ell,b}) (1 - n_{\ell+1,a}) - n_{\ell,b} n_{\ell+1,a}]$. While $I^\text{f}_\text{rr}$ quantifies the skin effect to the left in the single-particle scenario, $I^\text{k}_\text{rr}$ reveals the many-body skin effect to the right associated with the kink. Figure~\ref{fig:fig3}(c) displays $I^\text{f}_\text{rr}$ and $I^\text{k}_\text{rr}$ versus $V$, showing the DM-nCDW\textsubscript{k} transition at $t_1 = 1.2$. At $V=0$, fermions are tightly bound within each local resonance bond in DM, generating a relatively weak skin effect. As $V$ is switched on and increased, $I^\text{f}_\text{rr}$ initially decreases and then increases again, but reaches its maximum just to the right of the transition. As a many-body effect, $I^\text{k}_\text{rr}$ increases monotonically from DM to nCDW\textsubscript{k} with increasing $V$ and is generally an order of magnitude larger than $I^\text{f}_\text{rr}$, making it easier to detect experimentally. In addition, one can see that the excitation is gapped in the DM phase, but becomes gapless in the nCDW\textsubscript{k} phase. The presence of skin effects in both DM and nCDW\textsubscript{k} leads to a direction-dependent decaying behavior in the single-particle correlation function $\mathcal{C} (x) = \braket{c^\dag_{N - x + 1} c^{\phantom{\dag}}_{N + x + 1}}_\text{lr}$, as illustrated in Fig.~\ref{fig:fig3}(d).

\begin{figure}[!t]
\includegraphics[width=\columnwidth]{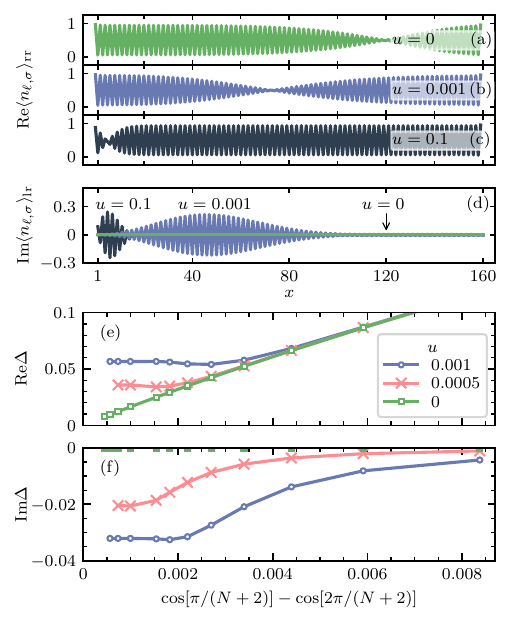}
\caption{\label{fig:fig4}(Color online) At $t_1=2$ and $V=10$, real parts of $\braket{n_{\ell,\sigma}}_\text{rr}$ in (a)-(c) and imaginary parts of $\braket{n_{\ell,\sigma}}_\text{lr}$ in (d) for a chain with $80$ unit cells, where the kink being sensitive to $u$ migrates towards the left as the chemical potential $u$ increases. (e)-(f): Real and imaginary parts of the gap $\Delta$ change with $N \leq 160$, non-vanishing in TDL for $u > 0$.}
\end{figure}

\textit{Localization of a kink at $u \ne 0$}.---Figure~\ref{fig:fig4} presents the complex fermion density in (a)-(d) and the energy gap in (e), (f). The skin effect of the kink for $u=0$ is shown with $\braket{n_{\ell,\sigma}}_\text{rr}$ in (a). For a finite $u$, one has an effective chemical potential of $\mu_\text{eff}=u(1-i)(2\ell-N)$~\cite{SM}. As $u$ grows, the skin effect of the kink on the right side gradually weakens, leaving the kink localized on the left side of the chain [see (b), (c)]. It is interesting to see in (d) that the imaginary part $\text{Im} \braket{n_{\ell,\sigma}}_\text{lr}$ forms a wave packet for the kink, and its size is essentially governed by $u$. Moreover,  the gap remains finite in the thermodynamic limit (TDL) for non-zero $u$ to characterize the localization of the kink, which can be found for both $\text{Re}\Delta$ and $\text{Im} \Delta$ in (e) and (f).

\textit{Summary and discussion}.---A biorthonormal-block density-matrix renormalization group algorithm is proposed for studying complex properties of non-Hermitian many-body systems, such as the spectrum, the energy gap, and other relevant observables. A block-diagonal form of the non-Hermitian reduced density matrix is conceptually introduced to partition the Hilbert spaces for each of semi-chains into the saved and discarded parts through a new truncation criterion. A two-step approach and the redundant degrees of freedom are implemented to construct the optimized saved space for the renormalization of operators. Numerical stability and efficiency are achieved by effectively reducing the condition number through additional skills and rescaling techniques. Accurate calculations are shown for system sizes that allow extrapolation to TDL with benchmarks provided for six models~\cite{SM}. It is straightforward to apply the algorithm to Liouvillian problems.

As applied to an interacting fermionic SSH model with nonreciprocal hoppings and a staggered complex chemical potential, a ground-state phase diagram is established with finding the nCDW and nCDW\textsubscript{k} phases, as well as a CDW-EP line. In particular, a new kind of skin effect emerges in connection with the dynamics of a kink, which gives rise to gapless excitations but becomes gapped in the presence of the chemical potential. The skin effect of the kink is an order of magnitude more pronounced than that of the fermion number, greatly enhancing the experimental visibility of non-Hermitian many-body physics.

\begin{acknowledgments}
We are grateful to Tao Xiang, Ulrich Schollw\"{o}ck, Ian Affleck, Walter Hofstetter, Zheng-Cheng Gu, Shi-Liang Zhu, and Shuo Yang for fruitful discussions. This work is supported by grants: MOST 2022YFA1402700, NSFC 12174020,  NSFC 12088101, NFSC 11974244, and NSAF U1930402. Computational resources from Tianhe-2JK at the Beijing Computational Science Research Center and Quantum Many-body-{\rm I} cluster at SPA, Shanghai Jiaotong University are also highly appreciated.
\end{acknowledgments}

\bibliography{ref}

\title{End Matter}

\maketitle

\textit{End~Matter~A:~Algorithm for diagonalization}.\label{app:diag}---To solve the eigenproblem of a non-Hermitian sparse matrix, e.g., the Hamiltonian for the superblock in bbDMRG, we utilize an advanced two-sided Krylov-Schur-restarted Arnoldi (TS-KSRA) algorithm~\cite{Zwaan_2017}, which combines the advantages of the early two-sided Arnoldi~\cite{Ruhe_1983} method and the Krylov-Schur restart technique~\cite{Stewart_2002}.
By employing a generalized Krylov decomposition that connects two distinct orthonormal basis sets for the space and its dual, it improves stability and accuracy~\cite{Stewart_2002}. Moreover, compared to the implicit QR decomposition used in ARPACK~\cite{lehoucq1998arpack}, TS-KSRA can efficiently anchor the converged Ritz vectors in the iteration~\cite{Zwaan_2017}.
The algorithm facilitates the correct acquisition of the corresponding left and right eigenstates for a specific degenerate state in bbDMRG, which is a challenging and often unfeasible task for the early one-sided algorithms.

\renewcommand{\theequation}{B\arabic{equation}}
\setcounter{equation}{0}

\textit{End~Matter~B:~Biorthonormal-block representation}.\label{app:bioForm}---We consider a chain of $L = n_\triangleleft + n_\triangleright$ sites as composed of semi-chains $\triangleleft$ and $\triangleright$, which contain $n_\triangleleft$ and $n_\triangleright$ sites, respectively. The left eigenstate $\bra{\bar{\Psi}}$ and the corresponding right eigenstate $\ket{\Psi}$ can be expressed as
\begin{eqnarray}
\label{eq:decompose}
\begin{split}
       \bra{\bar{\Psi}} \!=\! \sum_{\tau_\triangleleft,\tau_\triangleright} \bra{\tau_\triangleleft} \left(\bar{Y} \bar{\Theta}
	\bar{Z}\right)_{\tau_\triangleleft,\tau_\triangleright} \bra{\tau_\triangleright}
       = \begin{tikzpicture}[baseline=(X.base),every node/.style={scale=0.82},scale=.55]
		\draw[rounded corners] (1, 2) rectangle (2, 1); \draw (1.5, 1.5) node (X) {$\bar{Y}$}; \draw (2, 1.5) -- (3.3, 1.5); \draw (1.5, 1) -- (1.5, 0.5); \draw (2.5, 1.8) node (Y) {$\alpha$}; \draw (1.5, 0.2) node (Y) {$\tau_\triangleleft$};
		\node[rectangle,draw,yscale=1,rotate=45,minimum size=1ex] at (3.5, 1.5) {}; \node at (3.5, 2.1) {$\bar{\Theta}$};
		\draw (3.7, 1.5) -- (5, 1.5); \draw[rounded corners] (5, 2) rectangle (6, 1); \draw (5.5, 1.5) node {$\bar{Z}$}; \draw (5.5, 1) -- (5.5, 0.5); \draw (4.5, 1.8) node (Y) {$\alpha'$}; \draw (5.5, 0.2) node (Y) {$\tau_\triangleright$};
	\end{tikzpicture}\, ,\\
       \ket{\Psi} \!=\! \sum_{\tau_\triangleleft,\tau_\triangleright} \ket{\tau_\triangleleft} \left(Y \Theta
	Z\right)_{\tau_\triangleleft,\tau_\triangleright} \ket{\tau_\triangleright}
	= \begin{tikzpicture}[baseline=(X.base),every node/.style={scale=0.82},scale=.55]
		\draw[rounded corners] (1, 2) rectangle (2, 1); \draw (1.5, 1.5) node (X) {$Y$}; \draw (2, 1.5) -- (3.3, 1.5); \draw (1.5, 2) -- (1.5, 2.5); \draw (2.5, 1.2) node (Y) {$\alpha$}; \draw (1.5, 2.8) node (Y) {$\tau_\triangleleft$};
		\node[rectangle,draw,yscale=1,rotate=45,minimum size=1ex] at (3.5,1.5) {}; \node at (3.5, 0.9) {$\Theta$};
		\draw (3.7, 1.5) -- (5, 1.5); \draw[rounded corners] (5, 2) rectangle (6, 1); \draw (5.5, 1.5) node {$Z$}; \draw (5.5, 2) -- (5.5, 2.5); \draw (4.5, 1.2) node (Y) {$\alpha'$}; \draw (5.5, 2.8) node (Y) {$\tau_\triangleright$};
	\end{tikzpicture}\, ,
\end{split}
\end{eqnarray}
where $\tau_{\triangleleft,\triangleright}$ ($\alpha$, $\alpha^\prime$) represent physical (bond) indices for semi-chains. Bare bases $\ket{\tau_{\triangleleft,\triangleright}}$ are products of local physical bases $\ket{\tau_\ell}$, i.e., $\ket{\tau_\triangleleft} = \otimes_{\ell \in \triangleleft} \ket{\tau_\ell}$ and $\ket{\tau_\triangleright} = \otimes_{\ell \in \triangleright} \ket{\tau_\ell}$. Similarly, $\bra{\tau_\triangleleft} = \otimes_{\ell \in \triangleleft} \bra{\tau_\ell}$ and $\bra{\tau_\triangleright} = \otimes_{\ell \in \triangleright} \bra{\tau_\ell}$. Transformation matrices $Y$, $\bar{Y}$, $Z$, and $\bar{Z}$ fulfill the left and right biorthonormalization conditions (LBC and RBC), as follows by
\begin{eqnarray}\label{eq:LRBC}
	\bar{Y} Y \ = \
	\begin{tikzpicture}[baseline=(X.base),every node/.style={scale=0.82},scale=.55]
		\node (X) at (0.5, 2) {\quad};
                  \draw[rounded corners] (0.5,2) rectangle (1.5,1); \draw[rounded corners] (0.5,3.5) rectangle (1.5,2.5); \draw (1, 1.5) node {$Y$}; \draw (1, 3) node {$\bar{Y}$};
                  \draw (1, 2) -- (1, 2.5); \draw (1.5, 1.5) -- (2, 1.5); \draw (1.5, 3) -- (2, 3); 
	\end{tikzpicture}\ = \ \mathbb{1}\ , \quad \bar{Z} Z\ = \
	\begin{tikzpicture}[baseline=(X.base),every node/.style={scale=0.82},scale=.55]
		\node (X) at (0.5, 2) {};
		\draw (0.5, 1.5) -- (1, 1.5); \draw (0.5, 3) -- (1, 3); \draw (1.5,2) -- (1.5,2.5);
		\draw[rounded corners] (1,2) rectangle (2,1); \draw[rounded corners] (1,3.5) rectangle (2,2.5); \draw (1.5,1.5) node {$Z$}; \draw (1.5,3) node {$\bar{Z}$};
	\end{tikzpicture}\ = \ \mathbb{1}\, .
\end{eqnarray}
In bbDMRG, explicitly discussed in the main text, certain transformed bases are removed through truncation, so $\bar{Y}$ ($\bar{Z}$) is not the inverse matrix of $Y$ ($Z$). In contrast to the matrix-product-state representation~\cite{schollwock2011}, neither matrices $\bar{\Theta}$ nor $\Theta$ are generally diagonal.

Recursively performing the decomposition~\eqref{eq:decompose} results in a nested form of the eigenstates. Specifically,
\begin{eqnarray}
\label{eq:multidecomposition}
\begin{split}
	\bra{\bar{\Psi}} &= \sum_{\tau_\triangleleft,\tau_\triangleright} \bra{\tau_\triangleleft} \left(\bar{Y}_1 \cdots \bar{Y}_{n_\triangleleft} \bar{\Theta}
	\bar{Z}_{n_\triangleright} \cdots \bar{Z}_1\right)_{\tau_\triangleleft,\tau_\triangleright} \bra{\tau_\triangleright} \\
	 &= \begin{tikzpicture}[baseline=(X.base),every node/.style={scale=0.82},scale=.55]
		\draw[rounded corners] (0, 2) rectangle (1, 1); \draw (0.5, 1.5) node (X) {$\bar{Y}_1$}; \draw (0.5, 1) -- (0.5, 0.5); \draw (1, 1.5) -- (1.5, 1.5); 
		\draw (0.5, 0.2) node (Y) {$\tau_1$};
		\draw (2, 1.5) node {$\cdots$}; \draw (2.5, 1.5) -- (3, 1.5);
		\draw[rounded corners] (3, 2) rectangle (4, 1); \node at (3.5, 1.5) {$\bar{Y}_{n_\triangleleft}$}; \draw (3.5, 1) -- (3.5, 0.5); 
		\draw (4, 1.5) -- (4.8, 1.5); \draw (3.5, 0.2) node (Y) {$\tau_{n_\triangleleft}$};
		\node[rectangle,draw,yscale=1,rotate=45,minimum size=1ex] at (5, 1.5) {}; \node at (5, 2.1) {$\bar{\Theta}$};
		\draw (5.2, 1.5) -- (6, 1.5); \draw[rounded corners] (6, 2) rectangle (7, 1);
		\node at (6.5, 1.5) {$\bar{Z}_{n_\triangleright}$}; \draw (6.5, 1) -- (6.5, 0.5); \draw (7, 1.5) -- (7.5, 1.5);
		 \draw (6.5, 0.2) node (Y) {$\tau_{n_\triangleleft+1}$};
		\node at (8, 1.5) {$\cdots$};
		\draw (8.5,1.5) -- (9, 1.5); \draw[rounded corners] (9, 2) rectangle (10, 1); \node at (9.5, 1.5) {$\bar{Z}_1$}; \draw (9.5, 1) -- (9.5, 0.5); 
		\draw (9.5, 0.2) node (Y) {$\tau_L$};
	\end{tikzpicture}\, , \\ \ket{\Psi} &= \sum_{\tau_\triangleleft,\tau_\triangleright} \ket{\tau_\triangleleft} \left(Y_1 \cdots Y_{n_\triangleleft} \Theta Z_{n_\triangleright} \cdots Z_1\right)_{\tau_\triangleleft,\tau_\triangleright} \ket{\tau_\triangleright} \\
	&= \begin{tikzpicture}[baseline=(X.base),every node/.style={scale=0.82},scale=.55]
		\draw[rounded corners] (0, 2) rectangle (1, 1); \draw (0.5, 1.5) node (X) {$Y_1$}; \draw (0.5, 2) -- (0.5, 2.5); \draw (1, 1.5) -- (1.5, 1.5); \draw (0.5, 2.8) node (Y) {$\tau_1$};
		\draw (2, 1.5) node {$\cdots$}; \draw (2.5, 1.5) -- (3, 1.5);
		\draw[rounded corners] (3, 2) rectangle (4, 1); \node at (3.5, 1.5) {$Y_{n_\triangleleft}$}; \draw (3.5, 2) -- (3.5, 2.5); \draw (4, 1.5) -- (4.8, 1.5); \draw (3.5, 2.8) node (Y) {$\tau_{n_\triangleleft}$};
		\node[rectangle,draw,yscale=1,rotate=45,minimum size=1ex] at (5, 1.5) {}; \node at (5, 0.9) {$\Theta$};
		\draw (5.2, 1.5) -- (6, 1.5); \draw[rounded corners] (6, 2) rectangle (7, 1);
		\node at (6.5, 1.5) {$Z_{n_\triangleright}$}; \draw (6.5, 2) -- (6.5, 2.5); \draw (7, 1.5) -- (7.5, 1.5); \draw (6.5, 2.8) node (Y) {$\tau_{n_\triangleleft+1}$};
		\node at (8, 1.5) {$\cdots$};
		\draw (8.5,1.5) -- (9, 1.5); \draw[rounded corners] (9, 2) rectangle (10, 1); \node at (9.5, 1.5) {$Z_1$}; \draw (9.5, 2) -- (9.5, 2.5); \draw (9.5, 2.8) node (Y) {$\tau_L$};
	\end{tikzpicture}\, .
\end{split}
\end{eqnarray}
Each pair of transformation matrices $Y_\ell$ ($Z_\ell$) and $\bar{Y}_\ell$ ($\bar{Z}_\ell$) with $\ell \geq 2$ satisfies LBC and RBC, i.e.,
\begin{eqnarray}
	\bar{Y}_\ell Y_\ell \ = \
	\begin{tikzpicture}[baseline=(X.base),every node/.style={scale=0.82},scale=.55]
		\node (X) at (0.5, 2) {\quad};
		\draw[bend left=60] (0.5, 1.5) to (0.5, 3);
		\draw[rounded corners] (0.5,2) rectangle (1.5,1); \draw[rounded corners] (0.5,3.5) rectangle (1.5,2.5); 
		\draw (1, 1.5) node {$Y_\ell$}; \draw (1, 3) node {$\bar{Y}_\ell$};
		\draw (1, 2) -- (1, 2.5); \draw (1.5, 1.5) -- (2, 1.5); \draw (1.5, 3) -- (2, 3); 
	\end{tikzpicture}\ = \ \mathbb{1}\, , \ \ \bar{Z}_\ell Z_\ell \ = \
	\begin{tikzpicture}[baseline=(X.base),every node/.style={scale=0.82},scale=.55]
		\node (X) at (0.5, 2) {};
		\draw (0.5, 1.5) -- (1, 1.5); \draw (0.5, 3) -- (1, 3); \draw (1.5,2) -- (1.5,2.5);
		\draw[rounded corners] (1,2) rectangle (2,1); \draw[rounded corners] (1,3.5) rectangle (2,2.5); 
		\draw (1.5,1.5) node {$Z_\ell$}; \draw (1.5,3) node {$\bar{Z}_\ell$};
		\draw[bend right=60] (2, 1.5) to (2, 3); 
	\end{tikzpicture}\ = \ \mathbb{1}\, .
\end{eqnarray}
In particular, for $\ell = 1$, $Y_1$, $\bar{Y}_1$, $Z_1$ and $\bar{Z}_1$ satisfy Eq.~\eqref{eq:LRBC}. Using standard SVD technique~\cite{schollwock2011}, the biorthonormal-block representation of $\bra{\bar{\Psi}}$ and $\ket{\Psi}$ after multi-step decomposition can be readily transformed into a canonical form without sacrificing accuracy. This enables the measurement of the right-right expectation value of a physical observable.

Now, we briefly elucidate the flow at each bbDMRG step. For example, in the superblock structure ``B$\bullet\bullet$B"~\cite{White_1992, Peschel_1999, Schollwock_2005}, $\ket{y^\text{B}_\alpha}$ and $\bra{\bar{y}^\text{B}_\alpha}$ represent the biorthonormal bases for the left ``block", while $\ket{\tau^\bullet}$ and $\bra{\tau^\bullet}$ give the biorthonormal bases for the left ``point". We use these bases to construct those for the left semi-chain ``B$\bullet$", i.e., $\ket{y^\text{B}_\alpha} \otimes \ket{\tau^\bullet} \rightarrow \ket{y^0_\beta}$ and $\bra{\bar{y}^\text{B}_\alpha} \otimes \bra{\tau^\bullet} \rightarrow \bra{\bar{y}^0_\beta}$, which satisfy $\braket{\bar{y}^0_\beta \vert y^0_{\beta^\prime}} = \delta_{\beta,\beta^\prime}$. The index $\beta = 1$, $\cdots$, $d^\triangleleft$, where $d^\triangleleft$ denotes the dimension of spaces for the left semi-chain ``B$\bullet$". Similarly, we construct the biorthonormal bases $\ket{z^0_\beta}$ and $\bra{\bar{z}^0_\beta}$ for the right semi-chain ``$\bullet$B". The superblock Hamiltonian $H$ lives in the Hilbert spaces $\mathcal{H}$ and $\bar{\mathcal{H}}$ spanned by bases $\ket{y^0_\beta} \otimes \ket{z^0_{\beta'}}$ and $\bra{\bar{y}^0_\beta} \otimes \bra{\bar{z}^0_{\beta'}}$, respectively. Through the TS-KSRA algorithm as discussed in Appendix~A, the left and right eigenstates are simultaneously obtained and in form of
\begin{eqnarray}
\label{eq:dmrgphi}
\begin{split}
	\bra{\bar{\Psi}} &= \sum_{\beta,\beta'} \bra{\bar{y}^0_\beta} \bar{\Psi}_{\beta, \beta'} \bra{\bar{z}^0_{\beta'}} =
	\begin{tikzpicture}[baseline=(X.base),every node/.style={scale=0.82},scale=.55]
	 	\draw (3.8, 1.5) node (X) {$\beta$}; 
		\draw (4, 1.5) -- (4.8, 1.5);
		\node[rectangle,draw,yscale=1,rotate=45,minimum size=1ex] at (5, 1.5) {}; \node at (5, 2.1) {$\bar{\Psi}$};
		\draw (5.2, 1.5) -- (6, 1.5);
		\draw (6.2, 1.5) node (Y) {$\beta'$};
	\end{tikzpicture}\, , \\
	\ket{\Psi} &= \sum_{\beta, \beta'} \ket{y^0_\beta} \Psi_{\beta, \beta'} \ket{z^0_{\beta'}} =
	\begin{tikzpicture}[baseline=(X.base),every node/.style={scale=0.82},scale=.55]
	 	\draw (3.8, 1.5) node (X) {$\beta$}; 
		\draw (4, 1.5) -- (4.8, 1.5);
		\node[rectangle,draw,yscale=1,rotate=45,minimum size=1ex] at (5, 1.5) {}; \node at (5, 2.1) {$\Psi$};
		\draw (5.2, 1.5) -- (6, 1.5);
		\draw (6.2, 1.5) node (Y) {$\beta'$};
	\end{tikzpicture}\, .
\end{split}
\end{eqnarray}
A non-Hermitian reduced density matrix $\rho = \text{tr}_\triangleright \ketbra{\Psi} {\bar{\Psi}}$ resides in the spaces $\mathcal{H}^\triangleleft$ and $\bar{\mathcal{H}}^\triangleleft$ spanned by biorthonormal bases $\ket{y^0_\beta}$ and $\bra{\bar{y}^0_\beta}$. The normalization condition $\braket{\bar{\Psi} \vert \Psi} = 1$ ensures that $\mathrm{tr}_\triangleleft \rho = 1$. We can execute the block-diagonalization of $\rho$ following instructions stated in the main text and detailed techniques shown in Appendix~C, and obtain $\bar{Y}^{(\text{s})}$ and $Y^{(\text{s})}$ for ``B$\bullet$", as well as $\bar{Z}^{(\text{s})}$ and $Z^{(\text{s})}$ for ``$\bullet$B". After the successful construction of saved spaces, the transformed bases in the biorthonormal-block representation shown in Eq.~\eqref{eq:bWF} of the main text are defined as $\bra{\bar{y}_\alpha} = \sum_{\beta} \bar{Y}^{(\text{s})}_{\alpha,\beta} \bra{\bar{y}^0_\beta}$, $\bra{\bar{z}_{\alpha'}} = \sum_{\beta} \bar{Z}^{(\text{s})}_{\alpha',\beta} \bra{\bar{z}^0_\beta}$, $\ket{y_\alpha} = \sum_{\beta} \ket{y^0_\beta} Y^{(\text{s})}_{\beta,\alpha}$, and $\ket{z_{\alpha'}} = \sum_{\beta} \ket{z^0_\beta} Z^{(\text{s})}_{\beta,\alpha'}$, such that $\braket{\bar{y}_\alpha \vert y_{\alpha^\prime}} = \braket{\bar{z}_\alpha \vert z_{\alpha^\prime}} = \delta_{\alpha,\alpha^\prime}$. Thus the spectrum of $\rho$ is identical to that of $\Theta \bar{\Theta} = B^{(\text{s})}$, which validates the notion that $\Theta$ ($\bar{\Theta}$) represents the eigenstate $\ket{\Psi}$ ($\bra{\bar{\Psi}}$) under transformed bases. We then use the similarity transformation matrices $\bar{Y}^{(\text{s})}$, $Y^{(\text{s})}$, $\bar{Z}^{(\text{s})}$, and $Z^{(\text{s})}$ to complete the RG transformations of the Hamiltonians for semi-chains ``B$\bullet$" and ``$\bullet$B", as well as the necessary physical operators. Lastly, the left ``blocks" and ``point" become a new ``block", so do the right ``blocks" and ``point". We then turn to the next bbDMRG step.

\renewcommand{\theequation}{C\arabic{equation}}
\setcounter{equation}{0}

\textit{End~Matter~C:~Techniques for effectively reducing $\kappa$}.\label{app:techniques}---After obtaining the block diagonal form in Eq.~\eqref{eq:EqBlock} using a two-step approach, it is still encouraging to find appropriate $\eta$ for effectively reducing $\kappa$ according to redundancy~\eqref{eq:redundancy}.
To provide the recipes, we elucidate six relevant cases below.
Especially the instructions are given for the last two cases where we have to face the brute-force GEVD of $\rho$.

\textbf{(1)~Normal matrix}.
If $\rho$ is normal, which means that $[\rho,\ \rho^\dag] = 0$, both $\rho$ and $\rho^\dag$ can be diagonalized using the same unitary matrix $U$~\cite{Zhang_2017a}. In this case, we simply take $Y=U$ and $\bar{Y}=U^\dag$ so that $\bar{Y}Y=\mathbb{1}$ and $\kappa=1$ in bbDMRG, and $U$ can be obtained by diagonalizing the Hermitian matrix $(\rho + \rho^\dag) / 2$. The hybridization of $\rho$ and $\rho^\dagger$ allows for a seamless connection to the standard DMRG method for the Hermitian Hamiltonian.

\textbf{(2)~Null space}. A null space $\emptyset$, spanned by the left vectors $\bar{y}$ and the right vectors $y$ corresponding to zero eigenvalue, can not be gotten by the full diagonalization.

Analogous to the ``null'' subroutine in MATLAB, one may find $d$ orthonormal basis pairs
\begin{eqnarray}
	U^\emptyset =
	\begin{pmatrix}
		u^\rho_1 & \cdots & u^\rho_d
	\end{pmatrix}\, , \quad \quad V^\emptyset =
	\begin{pmatrix}
		v^\rho_1 & \cdots & v^\rho_d
	\end{pmatrix}\, ,
\end{eqnarray}
through the SVD of $\rho$, where $u^\rho$ and $v^\rho$ are left and right singular vectors for the singular value smaller than a threshold $\epsilon \ll 1$.
To do an eigenvalue decomposition of $V^{\emptyset\dag} U^\emptyset=CEC^{-1}$ using the standard LAPACK library, where $C$ is an invertible matrix and $E$ is a diagonal matrix, this yields biorthonormal vectors for RG transformations in bbDMRG
\begin{eqnarray}
y_{\alpha} = \sum_\alpha u_{\alpha'} C_{\alpha',\alpha} E^{-1/2}_{\alpha,\alpha}\, , \  \bar{y}_{\alpha} = \sum_{\alpha} E^{-1/2}_{\alpha,\alpha} C^{-1}_{\alpha,\alpha'} v^\dag_{\alpha'}\, .\ \ 
\end{eqnarray}
In practice, $\epsilon$ ranges from $10^{-14}$ to $10^{-11}$.

\textbf{(3)~Biorthonormalization}.
We can numerically improve the quality of the biorthonormalization conditions for the transformed bases by using the Gram-Schmidt-like algorithm~\cite{Kohaupt_2014}, which can reduce $\kappa$ greatly.

\textbf{(4)~Unitarization}.
As the model is close to EPs or other limits, ensuring numerical stability may become the most critical issue. To achieve this, we make the transformation matrices unitary, sacrificing some accuracy.

For the transformation matrix $Y^{(\mathrm{s})}$, one can obtain the SVD of $Y^{(\mathrm{s})}=U^{(\mathrm{s})} \Lambda^{(\mathrm{s})} V^{(\mathrm{s}) \dagger}$, where $U^{(\mathrm{s})}$ and $V^{(\mathrm{s})}$ are unitary matrices representing the orientations of the basis, and $\Lambda^{(\mathrm{s})}$ is an diagonal matrix that describes their amplitudes.
Since $\bar{Y}^{(\mathrm{s})} Y^{(\mathrm{s})}=\mathbb{1}$, we have $\bar{Y}^{(\mathrm{s})}=V^{(\mathrm{s})} {(\Lambda^{(\mathrm{s})})}^{-1} U^{(\mathrm{s}) \dagger}$ in general.
Now we discard the irrelevant amplitude $\Lambda^{(\mathrm{s})}$ so that the transformation matrices become unitary, i.e., $Y^{(s)'} = \bar{Y}^{(s)'\dagger} = U^{(s)} V^{(s)\dagger}$.
We note that this unitarization violates the biorthogonalization relation between the saved left vectors in $\bar{Y}^{(\mathrm{s})}$ and the discarded right vectors in $Y^{(\mathrm{d})}$, reducing truncation effectiveness.
In bbDMRG, if the energy change for $\triangleleft$ exceeds a threshold, we activate this operation.

\textbf{(5)~Jordan normal form}.
In the case of a non-diagonalizable matrix $\rho$, one can find at least a Jordan block with a rank of $r > 1$~\cite{SM}.
Standard libraries such as LAPACK and SciPy offer diagonalization subroutines that can typically be utilized to find the first right (generalized) eigenvector $y_1$ and the first left (generalized) eigenvector $\bar{y}_1$ for that Jordan block. If other pairs of generalized eigenvectors $y_\alpha$ and $\bar{y}_\alpha$ with $\alpha>1$ are required, the Jordan-chain algorithm~\cite{Wilkening_2007} can be used.
 
By applying the Jordan-chain recursion, one obtains a sequence of right generalized eigenvectors $y_\alpha$ such as 
\begin{eqnarray}
\label{eq::JordanChain}
	\left( \rho - \zeta \mathbb{1} \right) y_\alpha =  y_{\alpha-1}\, ,
\end{eqnarray}
for $\alpha=2$, $3$, $\dots$, $r$.
One may then form the transformation matrix $Y = (y_1 \cdots y_r)$ in bbDMRG. Using the Moore-Penrose inverse matrix~\cite{Zhang_2017a}, one obtains the other transformation matrix $\bar{Y}$ as follows:
\begin{eqnarray}
\bar{Y} = P Y^\dag\ ,
\end{eqnarray}
where $P$ is the inverse of $Y^\dag Y$~\cite{Zhang_2017a}.

\textbf{(6)~Degeneracy}.
It is possible to reduce $\kappa$ by combining $d$-fold degenerate eigenvector pairs of $y_\alpha$ and $\bar{y}_\alpha$ that share the same eigenvalue of $\rho$, which form $d \times d$ matrices
\begin{eqnarray}
	Y =
	\begin{pmatrix}
	y_1 & \cdots & y_d
	\end{pmatrix}\, , \quad \quad
	\bar{Y} =
	\begin{pmatrix}
	\bar{y}^\tran_1 & \cdots & \bar{y}^\tran_d
	\end{pmatrix}^\tran\, .
\end{eqnarray}
Then, we take a QR decomposition of $Y$, i.e., $Y = Q R$~\cite{Zhang_2017a}, where $Q$ is a unitary matrix and $R$ denotes an upper triangular matrix. The new transformation matrices are thus given by
\begin{eqnarray}
	Y^\prime = Q\, , \quad \quad \bar{Y}^\prime = R \bar{Y}\, ,
\end{eqnarray}
where $\kappa$ becomes smaller.

\end{document}


\title{\textbf{\large Supplemental Material for:\\ Density-matrix renormalization group algorithm for non-Hermitian systems}}

\author{Peigeng Zhong}
\affiliation{Beijing Computational Science Research Center, Beijing 100084, China}

\author{Wei Pan}
\affiliation{Beijing Computational Science Research Center, Beijing 100084, China}
\affiliation{Department of Physics, Beijing Normal University, Beijing, 100875, China}

\author{Haiqing Lin}
\email[]{haiqing0@csrc.ac.cn}
\affiliation{School of Physics, Zhejiang University, Hangzhou, 310058, China}
\affiliation{Beijing Computational Science Research Center, Beijing 100084, China}

\author{Xiaoqun Wang}
\email[]{xiaoqunwang@zju.edu.cn}
\affiliation{School of Physics, Zhejiang University, Hangzhou, 310058, China}

\author{Shijie Hu}
\email[]{shijiehu@csrc.ac.cn}
\affiliation{Beijing Computational Science Research Center, Beijing 100084, China}

\begin{abstract}
This supplemental material (SM) provides more details on the bbDMRG algorithm, including the Jordan normal form of the reduced density matrix, the analytical derivations for the upper error bound $\varepsilon_3$ 
and the effective model for dominant $V$, and benchmarks of different quantities for six different models. Major parts of SM verify the high precision, feasibility and capabilities of the bbDMRG algorithm from multiple aspects.
\end{abstract}

\maketitle

\renewcommand\thefigure{S\arabic{figure}}

\renewcommand{\theequation}{S\arabic{equation}}

\setcounter{figure}{0}

\section{I.~Jordan normal form of the reduced density matrix}
A non-Hermitian $\rho$ (any matrix in mathematics) can  be represented in a Jordan normal form $J$ which consists of $n_\text{J}$ Jordan blocks and can be obtained through a generalized eigenvalue decomposition (GEVD) as
\begin{eqnarray}
\label{eq:rhogevd}
\begin{split}
	\rho = Y J \bar{Y} &= \sum^{n_\text{J}}_{k=1} \sum_{\beta,\beta' = 1}^{r_k} y^{(k)}_\beta J^{(k)}_{\beta,\beta'} \bar{y}^{(k)}_{\beta'}\, ,
\end{split}
\end{eqnarray}
where $y^{(k)}_\beta$ and $\bar{y}^{(k)}_\beta$ are the right and left {\it generalized} eigenvectors of $\rho$, respectively. The Jordan block index $k$ varies from $1$ to $n_\text{J}$, and the inner index $\beta$ runs over the range $[1,\ r_k]$. The integer $r_k \ge 1$ denotes the dimension of the $k$-th Jordan block, which is equal to the rank of the block.
Consequently, the total dimension of the Hilbert space for the left semi-chain ($\triangleleft$) is given by $d^\triangleleft = \sum^{n_\text{J}}_{k=1} r_k$ when $\rho$ stands for a reduced density matrix for the left semi-chain.
In the $k$-th block, the matrix elements $J^{(k)}_{\beta,\beta} =\zeta_k$ and $J^{(k)}_{\beta,\beta+1}=1$, where $\zeta_k$ represents the equal eigenvalues, while all other elements are zero.

\begin{figure}[h]
\begin{center}
\includegraphics[width=0.7\textwidth]{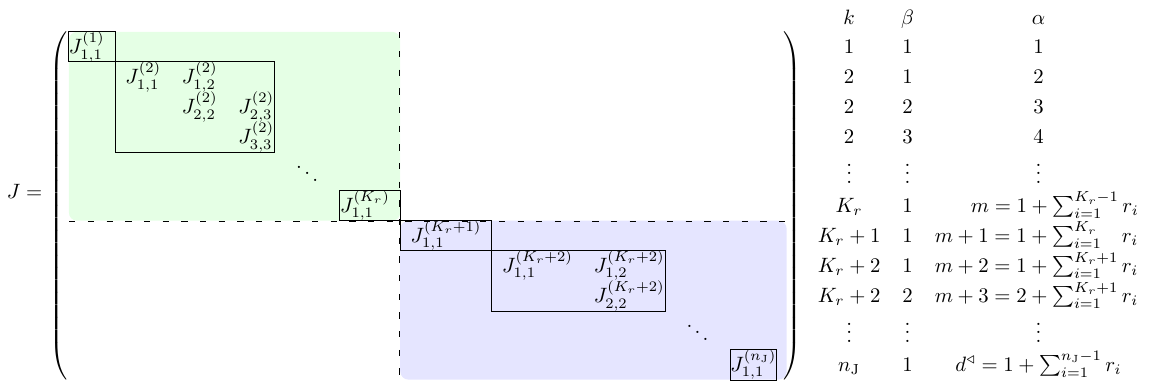}
\caption{\label{fig:figs1}
(Color online) Applying GEVD to the reduced density matrix $\rho$ for the left semi-chain yields a Jordan normal form $J$ consisting of $n_\text{J}$ Jordan blocks $J^{(k)}$, which are arranged in a descending order of the norms of eigenvalues $\lvert \zeta_k \rvert$. On the right-hand side, we demonstrate the binary index $(k, \beta)$ for the generalized eigenvector pair of $y^{(k)}_\beta$ and $\bar{y}^{(k)}_\beta$, labeled by a Jordan block index $k$ and an inner index $\beta$, as well as the equivalent single-element index $\alpha$. To construct the saved space (blocks with a green shaded background), we use the first $m$ pairs of the right eigenvectors $y_1$, $\cdots$, $y_m$ and the left eigenvectors $\bar{y}_1$, $\cdots$, $\bar{y}_m$, corresponding to the first $K_r$ Jordan blocks. The remaining $d^\triangleleft-m$ pairs of $y_{m+1}$, $\cdots$, $y_{d^\triangleleft}$ and $\bar{y}_{m+1}$, $\cdots$, $\bar{y}_{d^\triangleleft}$ (blocks with a blue shaded background) are discarded.}
\end{center}
\end{figure}

In the case of diagonalizable $\rho$, where $r_k=1$ for all $k$, we can write $\rho = \sum^{n_\text{J}}_{k=1} y^{(k)}_1 \zeta_k \bar{y}^{(k)}_1$ using an eigenvalue decomposition (EVD), where $y^{(k)}_1$ and $\bar{y}^{(k)}_1$ are the right and left eigenvectors of $\rho$.

For the sake of convenience in the upcoming discussions, we set up a convention that transforms a binary index $(k, \beta)$, which comprises a Jordan block index $k$ and an inner index $\beta$, into a regular single-element index $\alpha = \beta + \sum^{k-1}_{k'=1} r_{k'}$. A representative example depicting this convention is illustrated on the right side of Fig.~\ref{fig:figs1}.

When the dimension $m$ is small in the truncation procedure~\cite{White_1992, Peschel_1999, Schollwock_2005}, we only select the first $m$ (generalized) eigenvector pairs to optimize the construction of the saved space (s) and use the rest $d^\triangleleft-m$ pairs to build up the discarded space (d). 
The Jordan normal form in the saved space actually consists of the first $K_r$ Jordan blocks.
Thus, the dimension is also given by $m=\sum^{K_r}_{k=1} r_k$. $Y = (Y^{(\text{s})}\ Y^{(\text{d})})$ and $\bar{Y} = (\bar{Y}^{(\text{s})\tran}\ \bar{Y}^{(\text{d})\tran})^\tran$ represent two $d^\triangleleft \times d^\triangleleft$ transformation matrices, respectively, where all components are given by
\begin{eqnarray}
\label{eq:transformationmatrices}
	Y^{(\text{s})} &=
	\begin{pmatrix}
		y_1 & \cdots & y_m
	\end{pmatrix}\, , \quad Y^{(\text{d})} =
	\begin{pmatrix}
		y_{m+1} & \cdots & y_{d^\triangleleft}
	\end{pmatrix}\, , \quad \bar{Y}^{(\text{s})} =
	\begin{pmatrix}
		\bar{y}^\tran_1 & \cdots & \bar{y}^\tran_m
	\end{pmatrix}^\tran\, , \quad \bar{Y}^{(\text{d})} =
	\begin{pmatrix}
		\bar{y}^\tran_{m+1} & \cdots & \bar{y}^\tran_{d^\triangleleft}
	\end{pmatrix}^\tran\, .
\end{eqnarray}
In Eq.~\eqref{eq:transformationmatrices}, $y_\alpha$ is a $d^\triangleleft$-dimensional column vector, i.e., a $d^\triangleleft \times 1$ matrix, and $\bar{y}_\alpha$ is a $1 \times d^\triangleleft$ matrix, so $Y^{(\text{s})}$ is a $d^\triangleleft \times m$ matrix, $Y^{(\text{d})}$ is a $d^\triangleleft \times (d^\triangleleft-m)$ matrix, $\bar{Y}^{(\text{s})}$ is an $m \times d^\triangleleft$ matrix, and $\bar{Y}^{(\text{d})}$ is a $(d^\triangleleft-m) \times d^\triangleleft$ matrix.

\section{II.~Upper error bound $\epsilon_3$}

When the space $\bar{\mathcal{H}}$ is no longer the adjoint of the space $\mathcal{H}$ for a given $\rho$ so that the transformed bases for them are defined by $Y$ and $\bar{Y}$, respectively. The upper error bound
\begin{eqnarray}
\varepsilon_2 = a \left\Vert \rho - \rho^{(\text{s})} \right\Vert_2 = a \left\Vert \rho^{(\text{d})} \right\Vert_2 = a \left\Vert Y B^{(\text{d})} \bar{Y} \right\Vert_2
\end{eqnarray}
can be further analyzed to reach its minimization, where the saved part $\rho^{(\text{s})}$ is referred to as the \textit{structured} low-rank approximation of $\rho$~\cite{Chu_2003}.
The term ``structured" means that the similarity transformation required by the RG procedure for both the Hilbert space and its dual, introduces additional constraints on the two originally-independent unitary transformation matrices $U$ and $V$ in the conventional low-rank approximation.
These constraints prevent the structured low-rank approximation, viewed as a nonlinear multi-parameter optimization problem, from being converted into a linear optimization problem in general cases. Mathematically, it can be proven to be a complex \textit{non-convex optimization} problem, rendering the use of these $U$ and $V$ to construct the similarity transformation matrices impractical.
To handle the  problem, the Cauchy-Schwarz inequality allows us to introduce a larger UB $\varepsilon_3$ 
\begin{eqnarray}
  \begin{split}
  \varepsilon_2   \le a \left\Vert Y\right\Vert_2 \left\Vert B^{(\text{d})} \right\Vert_2 \left\Vert \bar{Y} \right\Vert_2\ 
=a \kappa \left\Vert B^{(\text{d})} \right\Vert_2 \propto a \kappa w_{m+1} = \varepsilon_3 \, .
\end{split}
\end{eqnarray}
where $\kappa = \Vert Y \Vert_2 \Vert \bar{Y}\Vert_2$ is called the condition number, measuring the strength on 
the deviation of the non-Hermitianity on $\rho$ from the optimal approximation of all possible Hermitian ones, or the deviation of the transformation matrix away from the optimal unitary one. While one finds that $\kappa=1$ for the unitary case, it is often  large and can become $\kappa \gg 1$ incredibly for the non-Hermitian cases, which often leads to severe numerical instability in the RG procedure.

For the $k$-th Jordan block $J^{(k)}$, we get the following bounds
\begin{eqnarray}
	\vert \zeta_k \vert \le \left\Vert J^{(k)} \right\Vert_2 \le \vert \zeta_k \vert + 1\, , 
\end{eqnarray}
where the left and right equalities in the above inequality hold for $r_k=1$ and $\infty$, respectively.
This weight $\left\Vert J^{(k)} \right\Vert_2$, similar to the weight $w_k$ defined in the main text,  enables us to assess the importance of the (generalized) eigenvectors corresponding to the $k$-th Jordan blocks, in order to decide whether to keep or discard these bases in the RG procedure.

\section{III.~Effective model for dominant $V$}

The Hamiltonian for the interacting fermionic SSH model defined in Eq.~(5) of the main text, can be divided into two parts, i.e., $H = H_0 + H_1$, where
\begin{eqnarray}\label{eq:nhham}
	H_0 = V \sum^N_{\ell=1} n^{\phantom{\dag}}_{\ell,a} n^{\phantom{\dag}}_{\ell,b} + V \sum^{N-1}_{\ell=1} n^{\phantom{\dag}}_{\ell,b} n^{\phantom{\dag}}_{\ell+1,a}\, ,
\end{eqnarray}
and $H_1$ contains other terms.

When $H_0$ is dominant due to large $V$, all $N$ fermions are either on the sublattice-$b$ for nCDW-$1$, or on the sublattice-$a$ for nCDW-$2$.
In addition, the bases $\ket{s_\ell} = \ket{\cdots 1_{\ell,a} 0_{\ell,b} 0_{\ell + 1,a} 1_{\ell + 1,b} \cdots}$ for a kink between the unit cells $\ell$ and $(\ell + 1)$ are also allowed.
Specifically, we set up $\ket{s_0} \equiv \ket{\text{nCDW-}1}$ and $\ket{s_N} \equiv \ket{\text{nCDW-}2}$ for $\ell = 0$ and $N$, resepctively.
Thus, we take $N+1$ bases into account in total.
As a result, we can obtain an effective Hamiltonian $H_\text{eff} =  H_\text{HN} + H_\text{IMP} + H_\text{LCP}$, up to the second order~\cite{Takahashi_1977, Mila_2011},
where $H^{\phantom{\text{NN}}}_\text{HN} = H^\text{NN}_\text{HN} + H^\text{NNN}_\text{HN}$,
\begin{eqnarray}
\label{eq:effHamCom}
	\begin{split}
		&H^\text{NN}_\text{HN} = \sum^{N-1}_{\ell=0} \left( t_\text{R} \ketbra{s_\ell}{s_{\ell+1}} + t_\text{L} \ketbra{s_{\ell + 1}}{s_\ell} \right)\, ,\\
		&H^\text{NNN}_\text{HN} = - \frac{1}{V} \sum^{N-2}_{\ell=0} \left( t^2_\text{R} \ketbra{s_\ell}{s_{\ell+2}} + t^2_\text{L} \ketbra{s_{\ell+2}}{s_\ell} \right)\, ,\\
		&H_\text{IMP} = - \frac{t^2_1 + t^2_2 - \gamma^2}{V} \left( \ketbra{s_0}{s_0} + \ketbra{s_N}{s_N} \right) + \text{const.}\, ,\\
		&H_\text{LCP} = \sum^N_{\ell=0} \mu_\text{eff} \ketbra{s_\ell}{s_\ell}\, .
	\end{split}
\end{eqnarray}
$H_\text{HN}$ describes the generalized Hatano-Nelson (HN) model~\cite{Ashida_2020}, consisting of the nearest-neighboring (NN) hopping term $H^\text{NN}_\text{HN}$ and the next-nearest-neighboring (NNN) one $H^\text{NNN}_\text{HN}$. 
The impurity (IMP) part $H_\text{IMP}$ gives the states $\ket{s_0}$ and $\ket{s_N}$ a lower real chemical potential due to open boundary conditions.
Lastly, the staggered complex chemical potential in the model~(5) simplifies to a linear chemical potential (LCP) $\mu_\text{eff} = u (1 - i) (2\ell - N)$ in $H_\text{LCP}$, as a linear function of the position of the kink.

We first consider the leading-order term $H^\text{NN}_\text{HN}$ in Eq.~\eqref{eq:effHamCom}, which gives an energy dispersion of $\epsilon = 2 \sqrt{t^2_1 - \gamma^2} \cos p$ with the lattice spacing set to $1$. 
The momentum $p = m p_0$ is given with $p_0=\pi / (N + 2)$ and an integer $m=1$, $\cdots$, $N+1$. As shown in the inset of Fig.~3(b) of the main text, the ground-state energy of the nCDW\textsubscript{k} state is given by $e_0 = -2 \sqrt{t^2_1 - \gamma^2} \cos p_0$ at the momentum $p=\pi - p_0$. For the second lowest-energy state at the momentum $p=\pi - 2 p_0$, we get its energy $e_1 = -2 \sqrt{t^2_1 - \gamma^2} \cos(2p_0)$. So the gap for the lowest excitation is given by
\begin{eqnarray}
\label{GAP}
\Delta = e_1 - e_0 = \chi \left[ \cos p_0 - \cos\left( 2p_0 \right) \right]
\end{eqnarray}
with $\chi = \chi_0 = 2 \sqrt{t^2_1 - \gamma^2}$. To include the NNN hopping term $H^\text{NNN}_\text{HN}$ in Eq.~\eqref{eq:effHamCom}, the energy dispersion can be approximated as $\epsilon \approx 2 \sqrt{t^2_1 - \gamma^2} \cos p - 2 [(t^2_1 - \gamma^2) / V] \cos (2 p)$ in the limit of $N \rightarrow +\infty$. As a result, the prefactor $\chi$ is renormalized by $V$, that is,
\begin{eqnarray}
\label{CHIHO}
	\chi = \chi_0 + \frac{4 (t^2_1 - \gamma^2)}{V} \left[ \cos p_0 + \cos\left( 2p_0 \right) \right] \approx \chi_0 + \frac{8 (t^2_1 - \gamma^2)}{V} > \chi_0 \, .
\end{eqnarray}
When the impurity chemical potential term $H_\text{IMP}$ is introduced, the prefactor $\chi$ becomes larger. For the original model, defined in Eq.~(5) of the main text, the scaling of the gap is given numerically by bbDMRG and shown in Fig.~3(b) of the main text, which keeps consistent very well with Eq.~\eqref{CHIHO}.

%

\section{III.~Benchmarks}

\subsection{A.~Sweeping process}

\begin{figure}[h]
	\includegraphics[width=0.8\textwidth]{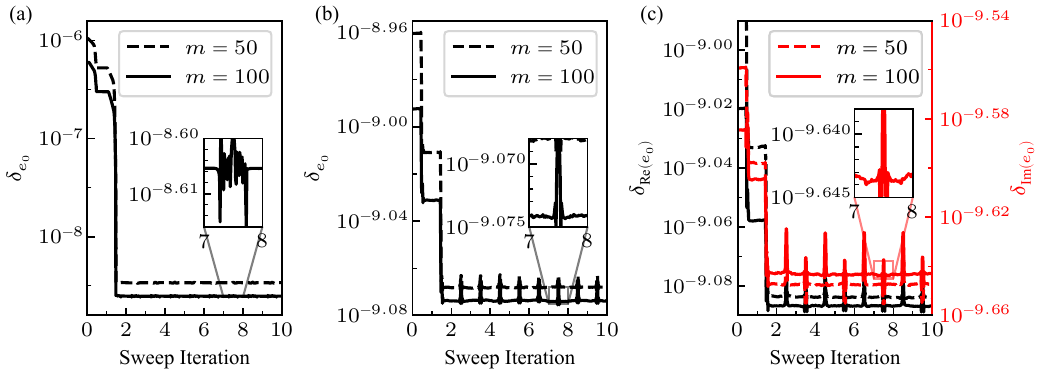}
	\caption{\label{fig:figs2}
	(Color online) Absolute errors of the target-state energy $e_0$ of the model~(5), used in the main text, as a function of the number of sweeps in the finite-chain bbDMRG algorithm, at $t_2=1$, $\gamma=0.1$, $V=0$ and $N=100$. The errors are defined as $\delta_{e_0} = \vert e_0 - e^\text{ext}_0 \vert$ for (a) $t_1=0.8$ in the topological insulator (TI), and (b) $t_1=1.5$ in the dimerized phase (DM), where the strength of the complex chemical potential $u=0$. While it can be evaluated separately as the error for the real part $\delta_{\text{Re}(e_0)} = \vert \text{Re}(e_0) - \text{Re}(e^\text{ext}_0) \vert$ (black) and the one for the imaginary part $\delta_{\text{Im}(e_0)} = \vert \text{Im}(e_0 - e^\text{ext}_0) \vert$ (red), respective, at (c) $t_1=1.5$ and $u=0.1$. To make a comparison, we choose two different dimensions $m=50$ (dashed line) and $100$ (solid line) in bbDMRG. Inset: the zoom-in plot in the $8$-th sweep iteration.}
\end{figure}

Figure~\ref{fig:figs2} shows how  the target state energy $e_0$ varies during the sweeping process for the model~(5) defined in the main text. 
For typical parameters within (a) the TI and (b) DM phase regions, $e_0$ remains real numbers and approaches the exact value with smaller errors in less than $2$ sweep iterations. 
In Fig~\ref{fig:figs2}(c), $e_0$ is a complex number when a complex chemical potential is introduced with the strength of $u=0.1$. It is remarkable that the errors for the real and imaginary parts decrease consistently during the finite-chain algorithm. 
Similar to Hermitian cases, the energy curves experience sudden spikes at the edges of the chain due to the significant growth of the truncation errors there.

\subsection{B.~Ground-state full variance}

\begin{figure}[h]
	\includegraphics[width=0.9\textwidth]{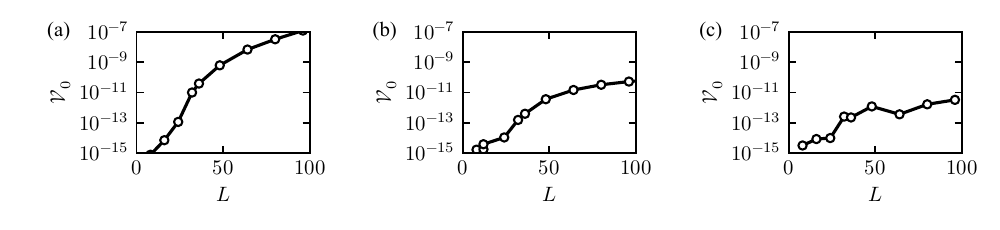}
	\caption{\label{fig:figs4}
	Ground-state full variance $\mathcal{V}_0$ as a function of the system size $L$ for both gapless and gapped phases.
	We consider three non-Hermitian quantum models under open boundary conditions at half-filling in the non-interacting limit $V=0$, including: (a) the HN model~\cite{Ashida_2020} at $J_\text{L}=1.1$ and $J_\text{R}=0.9$ (gapless), (b) the model~(5) used in the main text at $t_1=1.2$, $t_2=1$, $\gamma=0.1$ and $u=0$ (gapped), and (c) the SSH model with third-neighbor hoppings~\cite{Yao_2018} at $t_1=2$, $t_2=1$, $t_3=0.2$ and $\gamma=4/3$ (gapped).
	In bbDMRG, the dimension $m=100$ is used.}
\end{figure}

Figure~\ref{fig:figs4} shows the benchmarks of the ground-state full variance $\mathcal{V}_0 = \braket{(H-e_0)^2}_\text{lr}$~\cite{Hubig_2018} in three quantum models, corresponding to both gapped and gapless cases.
Here the word ``full”  emphasizes that we have already accounted for the effects caused by the truncated Hilbert space in the calculation of the variance.
For system sizes up to $L=100$, $\mathcal{V}_0$ remains remarkably low, consistently below $10^{-7}$ in both the gapless [Fig.~\ref{fig:figs4}(a)] and gapped phase regions [Figs.~\ref{fig:figs4}(b,c)].
In particular, for the SSH model with third-neighbor hoppings~\cite{Yao_2018}, $\mathcal{V}_0$ has an upper bound $10^{-11}$ for $L\le100$ [Fig.~\ref{fig:figs4}(c)].
These tests demonstrate the high quality of ground-state wave functions obtained from bbDMRG.

\subsection{C.~Ratios, fidelities and fidelity susceptibilities}
The quality of the ground-state wave functions can be further assessed by studying more quantities associated with the overlaps among the left and right eigenstates, as calculated through different ways discussed below.

Let us first consider the exact evaluations of those quantities in the non-interacting case, which will not only be used for comparison with the data obtained from bbDMRG in Table~\ref{table:ratio1}, but also be feasibly accessible for audiences to make their own comparisons.
For the simplest case, the non-Hermitian Hamiltonian $H$ has a Hermitian counterpart $H^{(\text{h})}=\nu H \nu^{-1}$ under the similarity transformation $\nu$.
For the SSH model~(5) used in the main text, the similarity transformation is given by $\nu = \prod^N_{\ell=1} g^{(\ell - 1) n^{\phantom{\dag}}_{\ell,a}} \prod^N_{\ell=1} g^{\ell n^{\phantom{\dag}}_{\ell,b}}$, where the coefficient $g=\sqrt{(t_1 + \gamma)/(t_1 - \gamma)} > 1$, and $n^{\phantom{\dag}}_{\ell,\sigma}$ denotes the particle-number operator for the fermion at site-($\ell,\sigma$).
For the HN model~\cite{Ashida_2020}, the similarity transformation is found to be $\nu = \prod^L_{\ell=1} g^{\ell n^{\phantom{\dag}}_\ell}$, with $g=\sqrt{J_\text{L}/J_\text{R}} > 1$, and $n^{\phantom{\dag}}_\ell$ denotes the particle-number operator for the fermion at site-$\ell$.
For both models, the many-body eigenstate can be represented as $\ket{\Psi^{(\text{h})}} = \otimes^N_{k=1} \ket{\phi^{(\text{h})}_k}$ which consists of $N$ normalized single-particle eigenstates $\ket{\phi^{(\text{h})}_k}$ at half-filling and satisfies the normalization condition $\braket{\Psi^{(\text{h})} \vert \Psi^{(\text{h})}} = 1$.
By applying the inverse of the similarity transformation matrix $\nu$ to $\ket{\Psi^{(\text{h})}}$, we obtain the right eigenstate $\ket{\Psi} = \nu^{-1} \ket{\Psi^{(\text{h})}} = \otimes^N_{k=1} \ket{\phi_k}$, where $\ket{\phi_k} = \nu^{-1} \ket{\phi^{(\text{h})}_k}$ may no longer be orthogonal.
So does the left eigenstate $\bra{\bar{\Psi}} = \bra{\Psi^{(\text{h})}} \nu = \otimes^N_{k=1} \bra{\bar{\phi}_k}$ with $\bra{\bar{\phi}_k} = \bra{\phi^{(\text{h})}_k} \nu$.
To reorthogonalize the right eigenstates, we employ QR decomposition, yielding that
\begin{eqnarray}\label{eq:QR}
\begin{pmatrix}
\ket{\phi_1} & \cdots & \ket{\phi_N}
\end{pmatrix}=
\begin{pmatrix}
\ket{\phi^\prime_1} & \cdots & \ket{\phi^\prime_N}
\end{pmatrix} Z\, ,
\end{eqnarray}
where $\ket{\phi^\prime_k}$ are orthonormal bases, and the diagonal elements of the upper-triangular matrix $Z$ provide the weights for $\ket{\phi^\prime_k}$.
Therefore, the norm of the right eigenstate is given by $\sqrt{\braket{\Psi \vert \Psi}} = \prod^N_{k=1} \vert Z_{kk} \vert$.
Similarly, the norm of the left eigenstate $\braket{\bar{\Psi} \vert \bar{\Psi}}$ can be evaluated.
Furthermore, the overlap of the left and right eigenstates is trivially equal to $1$, as $\braket{\bar{\Psi} \vert \Psi} = \braket{\Psi^{(\text{h})} \vert \Psi^{(\text{h})}} = 1$.
In the absence of the similarity transformation, we fully diagonalize the non-Hermitian Hamiltonian in single-particle bases.
A preexisting subroutine ``ZGEEV" from LAPACK is then used to obtain the normalized left and right eigenstates, ensuring that $\braket{\bar{\Psi}\vert\bar{\Psi}} = \braket{\Psi\vert\Psi} = 1$.

\begin{figure}[h]
	\includegraphics[width=0.8\textwidth]{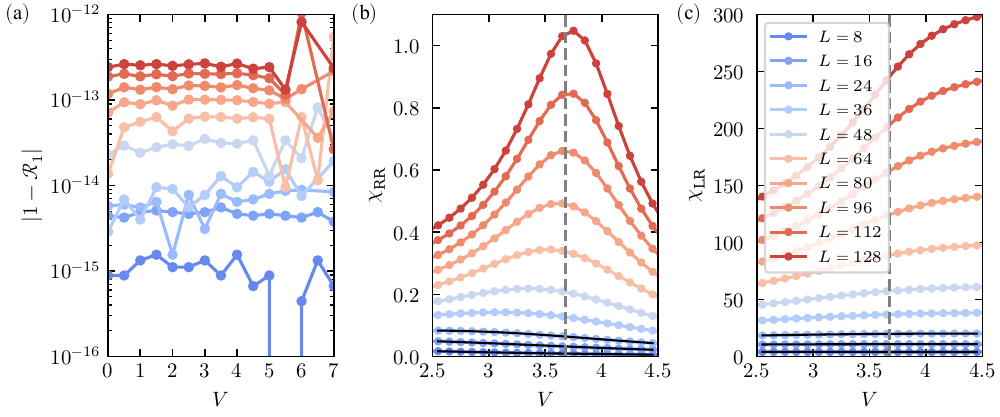}
	\caption{\label{fig:figs5}
	(Color online) (a) Ratio $\mathcal{R}_1$, (b) RR-fidelity susceptibility $\chi^{\phantom{\dag}}_\text{RR}$, and (c) LR-fidelity susceptibility $\chi^{\phantom{\dag}}_\text{LR}$ as a function of the repulsion strength $V$ for the model~(5) used in the main text at $t_1=1.2$, $t_2=1$, $\gamma=0.1$ and $u=0$. An interval of $\delta V=0.1$ is used. The exact values of the fidelity susceptibilities for $L = 2N \le24$ are marked by black solid lines. The dimension $m=100$ is used in bbDMRG.}
\end{figure}

Now we examine the quality of the ground-state wave functions for the SSH model~(5) used in the main text by evaluating the ratio $\mathcal{R}_1 = \vert \braket{\Psi \vert \nu^{-1} \vert \Psi^{(\text{h})}} \vert / \vert\braket{\Psi \vert \Psi}\vert$, where $\ket{\Psi}$ and $\ket{\Psi^{(\text{h})}}$ are obtained from bbDMRG and DMRG, respectively. Over a broad region $V\in[0,\, 7]$ for system sizes up to $L=2N=128$, the values of $\mathcal{R}_1$ keep close to the exact value $1$, with the largest deviation smaller than $10^{-12}$ [Fig.~\ref{fig:figs5}(a)].
This evidently shows that the quality of the ground-state right eigenstates obtained from bbDMRG can be as good as that calculated by applying the similarity transformation $\nu$ to the wave function given by the conventional DMRG calculation for the Hermitian counterpart of the SSH model~(5).

In addition, we define the right-right (RR) and left-right (LR) fidelities as
\begin{eqnarray}\label{eq:fidelity_t1}
\begin{split}
\mathcal{F}_\text{RR} = \frac{\vert \braket{\Psi (V + \delta V) \vert \Psi (V)} \vert}{\sqrt{\vert \braket{\Psi (V + \delta V) \vert \Psi (V + \delta V)} \braket{\Psi (V) \vert \Psi (V)} \vert}}\, ,\quad \mathcal{F}_\text{LR} = \frac{\vert \braket{\bar{\Psi} (V + \delta V) \vert \Psi (V)} \vert}{\sqrt{\vert \braket{\bar{\Psi} (V + \delta V) \vert \Psi (V + \delta V)} \braket{\bar{\Psi} (V) \vert \Psi (V)} \vert}}\, ,
\end{split}
\end{eqnarray}
which quantify the overlap between the ground-state wave functions for $V$ and $V + \delta V$, with other parameters kept fixed.
The corresponding fidelity susceptibilities are given by
\begin{eqnarray}\label{eq:fidelitysusceptibilities}
\chi^{\phantom{\dag}}_\text{RR} = - \frac{2\ln \mathcal{F}_\text{RR}}{(\delta V)^2}\, ,\qquad\chi^{\phantom{\dag}}_\text{LR} = - \frac{2\ln \mathcal{F}_\text{LR}}{(\delta V)^2}\, .
\end{eqnarray}
For small system sizes $L \le 24$ where the exact diagonalization calculation is available, the fidelity susceptibilities exhibit errors that are much smaller than the symbol size [Figs.~\ref{fig:figs5}(b,c)], 
indicating the high accuracy of bbDMRG results with only $m=100$ kept as compared to exact diagonalization results (black solid lines).
The RR-fidelity susceptibility $\chi^{\phantom{\dag}}_\text{RR}$ displays a peak near $V_c \approx 3.6$, which can be used accurately to signal a transition from the fermion skin effect to the kink skin effect.
It is remarkable that the transition points are identified in the phase diagram of the main text [Fig.~3(a)] by the peaks of $\chi^{\phantom{\dag}}_\text{RR}$, rather than the LR-fidelity susceptibility $\chi^{\phantom{\dag}}_\text{LR}$ which shows no singularities.

\begin{table}[t]
    \centering
    \begin{tabular}{|c|c|c|c|c||c|c|c|c|c|}
        \hline
        \multicolumn{10}{|c|}{The test for the ratio $\mathcal{R}_2$}\\
        \hline
	\multicolumn{10}{|c|}{(a) Noninteracting HN model~\cite{Ashida_2020} at $J_\text{L}=1.1$ and $J_\text{R}=0.9$ (gapless)}\\
	\hline
	$L$ & $\sqrt{\braket{\bar{\Psi} \vert \bar{\Psi}}}$ & $\sqrt{\braket{\Psi \vert \Psi}}$ & $\mathcal{R}_2$ (exact/\tcb{bbDMRG}) & $\delta_{\mathcal{R}_2}$ & $L$ & $\sqrt{\braket{\bar{\Psi} \vert \bar{\Psi}}}$ & $\sqrt{\braket{\Psi \vert \Psi}}$ & $\mathcal{R}_2$ (exact/\tcb{bbDMRG}) & $\delta_{\mathcal{R}_2}$ \\
	\hline
	\multirow{2}*{$4$} & \multirow{2}*{$1.36\times10^{0}$} & \multirow{2}*{$7.47\times10^{-1}$} & $0.982064128256513$ & \multirow{2}*{$0$} & \multirow{2}*{$8$} & \multirow{2}*{$4.21\times10^{0}$} & \multirow{2}*{$2.54\times10^{-1}$} & $0.936234394595596$ & \multirow{2}*{$2 \times 10^{-15}$} \\
	\cline{4-4} \cline{9-9}
    	& & & $\tcb{0.982064128256513}$ & & & & & $\tcb{0.936234394595598}$ & \\
	\hline
	\multirow{2}*{$12$} & \multirow{2}*{$2.94\times10^{1}$} & \multirow{2}*{$3.92\times 10^{-2}$} & $0.867738146603453$ & \multirow{2}*{$5 \times 10^{-13}$} & \multirow{2}*{$24$} & \multirow{2}*{$1.35\times10^{6}$} & \multirow{2}*{$1.27\times10^{-6}$} & $0.584242303504957$ & \multirow{2}*{$3 \times10^{-10}$} \\
	\cline{4-4} \cline{9-9}
	& & & $\tcb{0.867738146603925}$ & & & & & $\tcb{0.584242303785011}$ & \\
	\hline
	\multirow{2}*{$48$} & \multirow{2}*{$1.07\times10^{25}$} & \multirow{2}*{$7.51\times10^{-25}$} & $0.124821112014877$ & \multirow{2}*{$4 \times10^{-8}$} & \multirow{2}*{$64$} & \multirow{2}*{$5.25\times10^{44}$} & \multirow{2}*{$7.46\times10^{-44}$} & $0.025562426795309$ & \multirow{2}*{$1 \times10^{-7}$} \\
	\cline{4-4} \cline{9-9}
	& & & $\tcb{0.124821147943857}$ & & & & & $\tcb{0.025562532390592}$ & \\
	\hline
	\multirow{2}*{$80$} & \multirow{2}*{$1.22\times10^{70}$} & \multirow{2}*{$2.45\times10^{-68}$} & $0.003353571686365$ & \multirow{2}*{$4 \times10^{-7}$} & \multirow{2}*{$96$} & \multirow{2}*{$1.34\times10^{101}$} & \multirow{2}*{$2.65\times10^{-98}$} & $0.000281850568810$ & \multirow{2}*{$5 \times10^{-7}$} \\
	\cline{4-4} \cline{9-9}
	& & & $\tcb{0.003354011490601}$ & & & & & $\tcb{0.000282381151506}$ & \\
	\hline
        \hline
	\multicolumn{10}{|c|}{(b) Noninteracting SSH model~(5) used in the main text at $t_1=1.2$, $t_2=1$, $\gamma=0.1$ and $u=0$ (gapped)}\\
	\hline
	$L$ & $\sqrt{\braket{\bar{\Psi} \vert \bar{\Psi}}}$ & $\sqrt{\braket{\Psi \vert \Psi}}$ & $\mathcal{R}_2$ (exact/\tcb{bbDMRG}) & $\delta_{\mathcal{R}_2}$ & $L$ & $\sqrt{\braket{\bar{\Psi} \vert \bar{\Psi}}}$ & $\sqrt{\braket{\Psi \vert \Psi}}$ & $\mathcal{R}_2$ (exact/\tcb{bbDMRG}) & $\delta_{\mathcal{R}_2}$ \\
	\hline
	\multirow{2}*{$12$} & \multirow{2}*{$4.57\times10^{0}$} & \multirow{2}*{$2.26\times10^{-1}$} & $0.968754130774826$ & \multirow{2}*{$5\times10^{-15}$} & \multirow{2}*{$24$} & \multirow{2}*{$4.25\times10^{2}$} & \multirow{2}*{$2.54\times10^{-3}$} & $0.928261070962912$ & \multirow{2}*{$6\times10^{-11}$} \\
	\cline{4-4} \cline{9-9}
	& & & $\tcb{0.968754130774821}$ & & & & & $\tcb{0.928261071022878}$ & \\
	\hline
	\multirow{2}*{$48$} & \multirow{2}*{$3.04\times10^{10}$} & \multirow{2}*{$3.88\times10^{-11}$} & $0.848305991535367$ & \multirow{2}*{$9\times10^{-9}$} & \multirow{2}*{$64$} & \multirow{2}*{$4.19\times10^{18}$} & \multirow{2}*{$2.99\times10^{-19}$} & $0.798569358495656$ & \multirow{2}*{$4\times10^{-8}$} \\
	\cline{4-4} \cline{9-9}
	& & & $\tcb{0.848306000663948}$ & & & & & $\tcb{0.798569401736848}$ & \\
	\hline
	\multirow{2}*{$80$} & \multirow{2}*{$1.21\times10^{29}$} & \multirow{2}*{$1.10\times10^{-29}$} & $0.751735668280499$ & \multirow{2}*{$1\times10^{-7}$} & \multirow{2}*{$96$} & \multirow{2}*{$7.32\times10^{41}$} & \multirow{2}*{$1.93\times10^{-42}$} & $0.707647173212049$ & \multirow{2}*{$2\times10^{-7}$} \\
	\cline{4-4} \cline{9-9}
	& & & $\tcb{0.751735768839196}$ & & & & & $\tcb{0.707647336451347}$ & \\
	\hline
	\multirow{2}*{$112$} & \multirow{2}*{$9.29\times10^{56}$} & \multirow{2}*{$1.62\times10^{-57}$} & $0.666144246890851$ & \multirow{2}*{$2\times10^{-7}$} & \multirow{2}*{$128$} & \multirow{2}*{$2.47\times10^{74}$} & \multirow{2}*{$6.45\times10^{-75}$} & $0.627075410859138$ & \multirow{2}*{$3\times10^{-7}$} \\
	\cline{4-4} \cline{9-9}
	& & & $\tcb{0.666144470624514}$ & & & & & $\tcb{0.627075688720030}$ & \\
	\hline
        \hline
	\multicolumn{10}{|c|}{(c) Noninteracting SSH model with third-neighbor hoppings~\cite{Yao_2018} at $t_1=2$, $t_2=1$, $t_3=0.2$ and $\gamma=4/3$ (gapped)}\\
	\hline
	$L$ & $\sqrt{\braket{\bar{\Psi} \vert \bar{\Psi}}}$ & $\sqrt{\braket{\Psi \vert \Psi}}$ & $\mathcal{R}_2$ (exact/\tcb{bbDMRG}) & $\delta_{\mathcal{R}_2}$ & $L$ & $\sqrt{\braket{\bar{\Psi} \vert \bar{\Psi}}}$ & $\sqrt{\braket{\Psi \vert \Psi}}$ & $\mathcal{R}_2$ (exact/\tcb{bbDMRG}) & $\delta_{\mathcal{R}_2}$ \\
	\hline
	\multirow{2}*{$8$} & \multirow{2}*{$\slash$} & \multirow{2}*{$\slash$} & $0.710796248145485$ & \multirow{2}*{$5\times10^{-15}$} & \multirow{2}*{$16$} & \multirow{2}*{$\slash$} & \multirow{2}*{$\slash$} & $0.474160890425344$ & \multirow{2}*{$1\times10^{-13}$} \\
	\cline{4-4} \cline{9-9}
	& & & $\tcb{0.710796248145490}$ & & & & & $\tcb{0.474160890425482}$ & \\
	\hline
	\multirow{2}*{$24$} & \multirow{2}*{$\slash$} & \multirow{2}*{$\slash$} & $0.315826719123342$ & \multirow{2}*{$8\times10^{-10}$} & \multirow{2}*{$32$} & \multirow{2}*{$\slash$} & \multirow{2}*{$\slash$} & $0.210344134635943$ & \multirow{2}*{$1\times10^{-9}$} \\
	\cline{4-4} \cline{9-9}
	& & & $\tcb{0.315826719875463}$ & & & & & $\tcb{0.210344135811114}$ & \\
	\hline
	\multirow{2}*{$48$} & \multirow{2}*{$\slash$} & \multirow{2}*{$\slash$} & $0.093300595262410$ & \multirow{2}*{$3\times10^{-9}$} & \multirow{2}*{$64$} & \multirow{2}*{$\slash$} & \multirow{2}*{$\slash$} & $0.041384493501080$ & \multirow{2}*{$3\times10^{-9}$} \\
	\cline{4-4} \cline{9-9}
        & & & $\tcb{0.093300597812007}$ & & & & & $\tcb{0.041384496568260}$ & \\
	\hline
	\multirow{2}*{$80$} & \multirow{2}*{$\slash$} & \multirow{2}*{$\slash$} & $0.018356541065514$ & \multirow{2}*{$2\times10^{-9}$} & \multirow{2}*{$96$} & \multirow{2}*{$\slash$} & \multirow{2}*{$\slash$} & $0.008142242918445$ & \multirow{2}*{$1\times10^{-9}$} \\
	\cline{4-4} \cline{9-9}
        & & & $\tcb{0.018356542877411}$ & & & & & $\tcb{0.008142243929534}$ & \\
	\hline
    \end{tabular}
    \caption{Norms, the ratio $\mathcal{R}_2$ and the absolute error $\delta_{\mathcal{R}_2}$ for both gapped and gapless ground states of three models at half-filling. The values of $\mathcal{R}_2$ obtained from bbDMRG are highlighted in blue, 
    with a dimension of $m=200$ for (a), and $m=100$ for (b,c). In (c), due to the lack of a similarity transformation that can map the non-Hermitian Hamiltonian to its Hermitian counterpart, 
    we cannot obtain the exact estimates of the norms of the left and right eigenstates, which are marked with ``$\slash$".}
    \label{table:ratio1}
\end{table}

Furthermore, we also calculated  the norms $\sqrt{\braket{\bar{\Psi} \vert \bar{\Psi}}}$ and $\sqrt{\braket{\Psi \vert \Psi}}$, as well as the ratio $\mathcal{R}_2 = \vert \braket{\bar{\Psi} \vert \Psi} \vert/\sqrt{\vert \braket{\bar{\Psi} \vert \bar{\Psi}} \braket{\Psi \vert \Psi} \vert}$, 
for the ground states of three non-Hermitian models. As presented in Table~\ref{table:ratio1}, one can see that  bbDMRG results  of $\mathcal{R}_2 $ agree well with those available exact results,  and $\delta_{\mathcal{R}_2}$ remains very small for various sizes of the systems, which evidently exhibits the high accuracy of the bbDMRG calculations.
However, the norms of left and right eigenstates behave dramatically differently with respect to $L$, resulting in severe numerical challenges in early direct extensions of DMRG to non-Hermitian Hamiltonians, associated with the huge condition number $\kappa$. Further discussions on this can be found in the main text.

\begin{figure}[t]
	\includegraphics[width=0.7\textwidth]{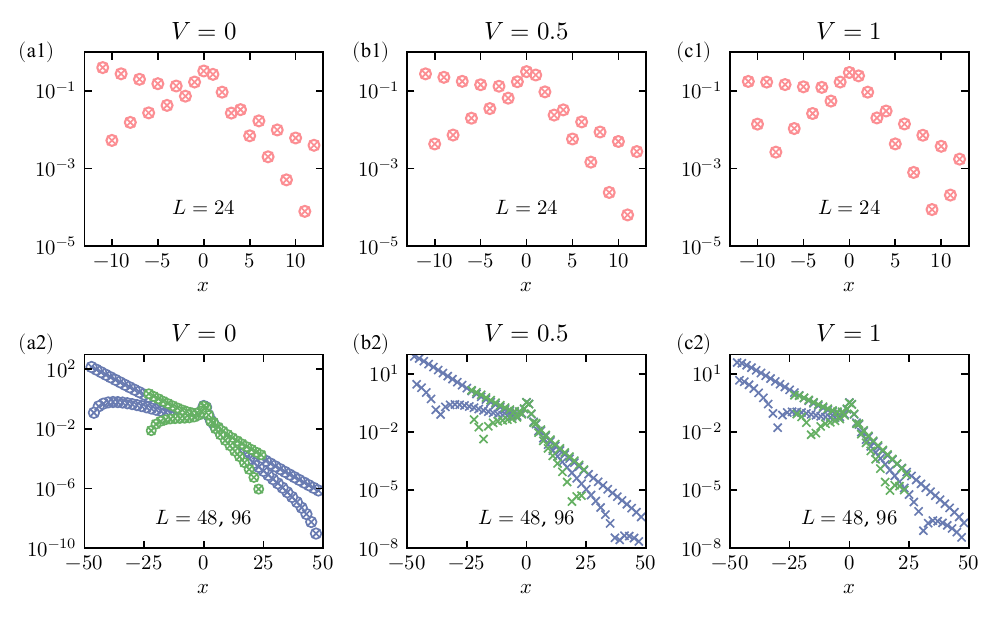}
	\caption{\label{fig:figs6_1}
	(Color online) Single-particle correlation function $\mathcal{C}(x)$ for the HN model~\cite{Ashida_2020} at $J_\text{L}=1.1$, $J_\text{R}=0.9$ and half-filling. We consider three cases: (a1,a2) non-interacting $V=0$, (b1,b2) with $V=0.5$, and (c1,c2) with $V=1$. Three typical system sizes are also considered: $L=24$ (red), $48$ (green) and $96$ (blue). Exact values are marked by $\circ$, while the data obtained from bbDMRG are denoted by $\times$. The dimension $m=200$ is used in bbDMRG.}
\end{figure}

\begin{figure}[b]
	\includegraphics[width=0.7\textwidth]{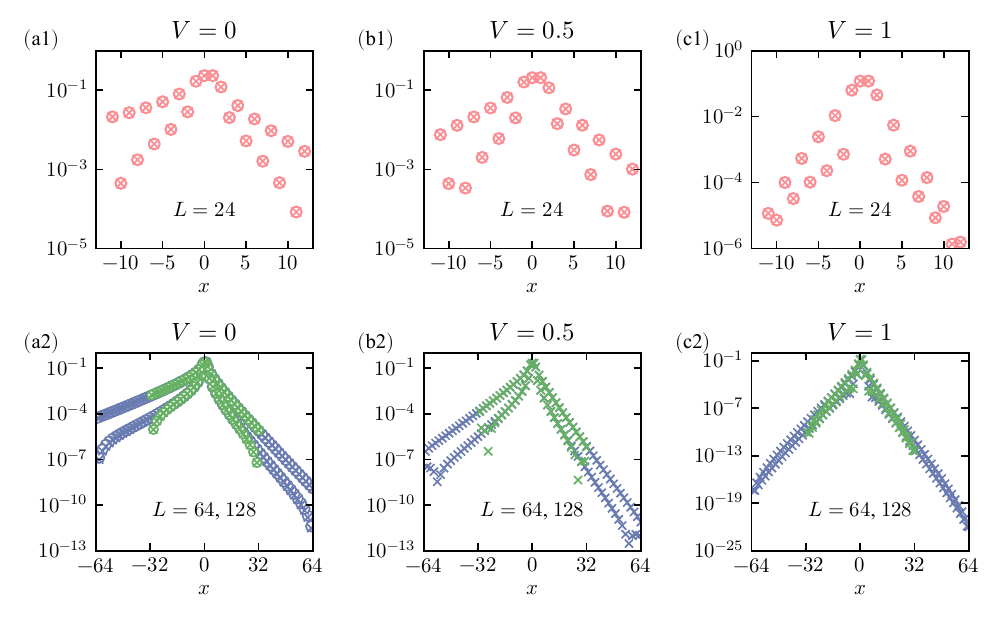}
	\caption{\label{fig:figs6_2}
	(Color online) Single-particle correlation function $\mathcal{C}(x)$ for the SSH model~(5) used in the main text at $t_1=1.2$, $t_2=1$, $\gamma=0.1$, $u=0$ and half-filling. We consider three cases: (a1,a2) $V=0$, (b1,b2) $1$, and (c1,c2) $6$. Three typical system sizes are also considered: $L=24$ (red), $64$ (green) and $128$ (blue). Exact values are marked by $\circ$, while the data obtained from bbDMRG are denoted by $\times$. The dimension $m=100$ is used in bbDMRG.}
\end{figure}

\subsection{D.~Single-particle correlation function}
\begin{figure}[t]
	\includegraphics[width=0.7\textwidth]{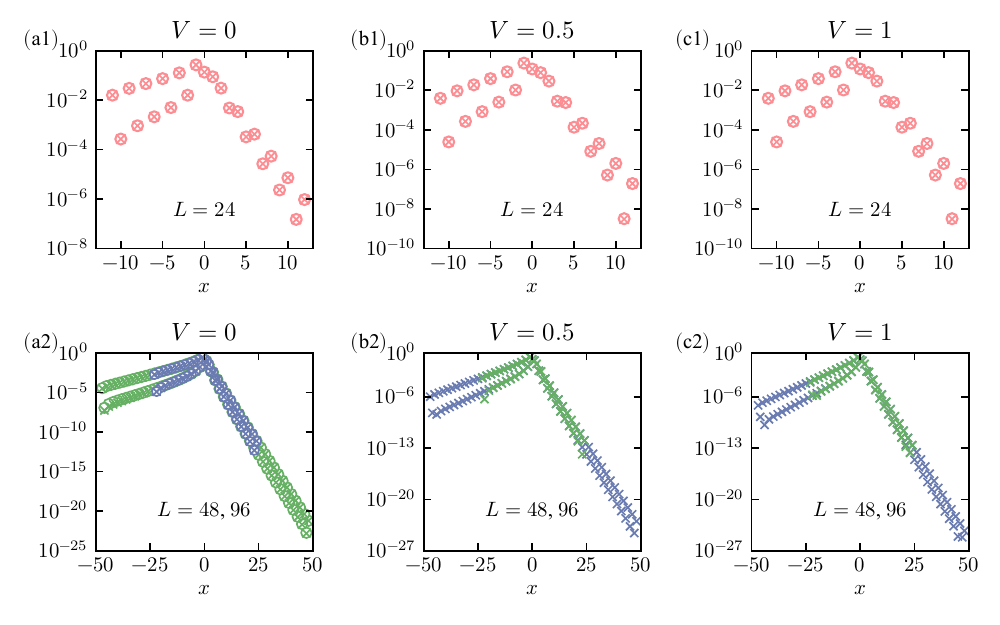}
	\caption{\label{fig:figs6_3}
	(Color online) Single-particle correlation function $\mathcal{C}(x)$ for the SSH model with third-neighbor hoppings~\cite{Yao_2018} at $t_1=2$, $t_2=1$, $t_3=0.2$, $\gamma=4/3$ and half-filling. We consider three cases: (a1,a2) $V=0$, (b1,b2) $0.5$, and (c1,c2) $1$. Three typical system sizes are also considered: $L=24$ (red), $48$ (green) and $96$ (blue). The values calculated from Eq.~\eqref{eq:SPCFExact} using $50$-digit floating-point numbers are marked by $\circ$, respectively, while the data obtained from bbDMRG are denoted by $\times$. The dimension $m=100$ is used in bbDMRG.}
\end{figure}

A single-particle correlation function (SPCF) is an appropriate quantity to reveal the quality of ground-state wave functions, as defined by
\begin{eqnarray}\label{eq:SPCF}
\mathcal{C}(x) = \braket{c^\dag_{L/2 - x + 1} c^{\phantom{\dag}}_{L/2 + x + 1}}_\text{lr} \propto e^{-2\left(\omega_0 x + \vert x \vert / \xi\right)}\, ,
\end{eqnarray}
which exhibits a direction-dependent exponentially-decaying behavior at long-distances.
The correlation length $\xi$ is associated with a finite bulk gap, while an extra prefactor $\omega_0$ of $x$ arises from non-Hermitian skin effects so that the SPCF depends on the sign of the displacement $x$.
In particular, for $x > 0$, the creation operator is located to the left of the annihilation operator in Eq.~\eqref{eq:SPCF}, introducing an additional decaying effect in the long-distance asymptotic behavior of SPCF.
Conversely, when $x < 0$, the long-range SPCF is enhanced by the skin effects, leading to a novel exponential growth, particularly evident in the HN model [Fig.~\ref{fig:figs6_1}].
We evaluate the exact estimate of SPCF for sites $x$ and $x^\prime$ with employing the formula
\begin{eqnarray}\label{eq:SPCFExact}
\braket{c^\dag_{x} c^{\phantom{\dag}}_{x^\prime}}_\text{lr} = \begin{pmatrix}
\braket{x \vert \phi_1} & \cdots & \braket{x \vert \phi_N}
\end{pmatrix}
\begin{pmatrix}
\braket{\bar{\phi}_1 \vert x^\prime} \\
\vdots \\
\braket{\bar{\phi}_N \vert x^\prime}
\end{pmatrix}\, ,
\end{eqnarray}
where $\ket{x}$ denotes the bases for the coordinate representation.
It is once more to notice that the data given in the tests [Figs.~\ref{fig:figs6_1}, \ref{fig:figs6_2}, and \ref{fig:figs6_3}], obtained from bbDMRG, match very well with the exact results for $L\le 24$ and the exact solution for larger sizes in the non-interacting cases.
The even-odd oscillation behavior of $\mathcal{C}(x)$ results from the alternating hopping terms in the model~(5) used in the main text.
Based on the accurate results for SPCF, one can expect the reliability of bbDMRG computations for other quantities defined by the ground-state expectations.

\begin{figure}[t]
	\includegraphics[width=0.8\textwidth]{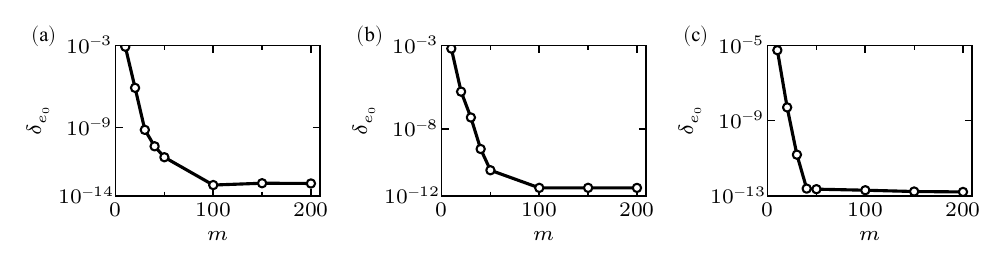}
	\caption{\label{fig:figs3}
	(Color online) Absolute errors $\delta_{e_0}$ of the ground-state energy $e_0$ as a function of the dimension $m$ used in bbDMRG after $6$ sweep iterations are applied at least. We consider three non-Hermitian quantum models under open boundary conditions, including (a) the $\mathcal{PT}$-symmetric SSH chain~\cite{Lieu_2018a} of $24$ sites at $v_1=v_2=\sin(\pi/8)$, $w_1=w_2=\cos(\pi/8)$, $u=0.3$, $V=2$ and half-filling, (b) the non-Hermitian spin-$1/2$ XXZ chain~\cite{Yamamoto_2022a} of $24$ sites at $J=1$ and $\Delta_\gamma=1.5+0.5i$, (c) the non-Hermitian Aubrey-Andr\'{e}-Harper (AAH) Bose-Hubbard chain~\cite{Zhang_2020} of $18$ sites at $J=V=1$, $\gamma=0.4$, $\alpha=1/3$, $\delta=2\pi/3$, $U=4$ and $1/3$-filling. In the Fock bases for each site, the maximum number of bosons is limited to $4$.}
\end{figure}

\subsection{E.~Other quantum models}

Figure~\ref{fig:figs3} provides additional benchmarks of the ground-state energy in other non-Hermitian models, including $\mathcal{PT}$-symmetric SSH chain of interacting fermions~\cite{Lieu_2018a}, non-Hermitian spin-$1/2$ XXZ chain~\cite{Yamamoto_2022a}, non-Hermitian Aubrey-Andr\'{e}-Harper (AAH) Bose-Hubbard chain~\cite{Zhang_2020} under open boundary conditions. It is worth noting that in the ordinary $\mathcal{PT}$-symmetric SSH chain studies in Ref.~\cite{Lieu_2018a}, we also add the repulsive interaction term. These results further demonstrate the efficiency of the bbDMRG algorithm, where one can see that the absolute errors of the ground-state energy consistently diminish very rapidly with increasing the dimension up to  $m=100$  and approach to the double-precision limit of floating-point numbers for all three models.

\bibliography{refsm}